\documentclass[journal,compsoc]{IEEEtran}
\usepackage{url}
\usepackage{enumitem}
\usepackage[colorlinks=true,linkcolor=blue,citecolor=blue,urlcolor=blue]{hyperref}
\usepackage{orcidlink}
\usepackage{xcolor}
\usepackage[ruled,linesnumbered]{algorithm2e} 

\SetAlFnt{\small}
\SetAlCapFnt{\small}
\SetAlCapNameFnt{\small}
\SetAlCapHSkip{0pt}
\SetKwInput{KwInput}{Input}
\SetKwInput{KwRequire}{Require}
\usepackage[cal=cm] {mathalfa}

\SetCommentSty{\normalfont\rmfamily\upshape}

\usepackage{amsmath}
\usepackage{caption}

\usepackage{multirow}
\usepackage{xspace}

\makeatletter
\DeclareRobustCommand\onedot{\futurelet\@let@token\@onedot}
\def\@onedot{\ifx\@let@token.\else.\null\fi\xspace}

\def\ie{\emph{i.e}\onedot}

\def\etal{\emph{et al}\onedot}
\makeatother

\usepackage{mathptmx} 
\usepackage{tabu}
\usepackage{booktabs}   
\usepackage{mwe}
\usepackage{listings}
 \lstdefinelanguage{json}{ basicstyle=\fontsize{7}{7.5}\selectfont\ttfamily, showstringspaces=false, breaklines=true, frame=lines, tabsize=2, keywordstyle=[1]\color{brown}, keywordstyle=[2]\color[rgb]{0,0.6,0}, captionpos=b, backgroundcolor=\color[RGB]{245,245,244}, morekeywords=[1]{selectIngredient, selectSkeleton, createRule, editRule, saveModel, loadModel, updatePivot, updatePosition, highlightIngredient, modifyColor, changeMode, elements, distance, collisionDetection,space, alignDirection,length,tweaking,std,curve,labeling}, morekeywords=[2]{Str, Int, vector3D, Bool,Double}, literate= *{:}{{{\color{gray}{:}}}}{1} {,}{{{\color{gray}{,}}}}{1} {\{}{{{\color{blue}{\{}}}}{1} {\}}{{{\color{blue}{\}}}}}{1} {[}{{{\color{blue}{[}}}}{1} {]}{{{\color{blue}{]}}}}{1}, }
\graphicspath{{figs/}{figures/}{pictures/}{images/}{./}}

\usepackage{tabu}                      
\usepackage{booktabs}                  
\usepackage{lipsum}                    
\usepackage{mwe}                       

\usepackage{mathptmx}                  

\begin{document}
\title{Chat Modeling: Interaction-Enhanced Agent Framework for Visualizing Literature-Grounded Biological Structures}
\author{Donggang Jia\orcidlink{0000-0002-1358-8718}, Yunhai Wang\orcidlink{0000-0003-0059-6580}, and Ivan Viola\orcidlink{0000-0003-4248-6574}
\thanks{Donggang Jia and Ivan Viola are with the Computer, Electrical and Mathematical Sciences and Engineering (CEMSE) Division, King Abdullah University of Science and Technology
(KAUST), Saudi Arabia. E-mail: \{donggang.jia\,$|$\,ivan.viola\}@kaust.edu.sa.}
\thanks{Yunhai Wang is with the School of Information, Renmin University of China, China.
   E-mail: wang.yh@ruc.edu.cn.}
\thanks{Ivan Viola and Yunhai Wang are co-corresponding authors.}
}

\markboth{Accepted to IEEE Transactions on Visualization and Computer Graphics (TVCG), 2026}%
{}

\maketitle
\begin{abstract}
Bioscientists frequently seek to visualize the biological systems they have empirically characterized and reported in the literature. Realizing such visualizations requires biological structure modeling, an inherently complex process that demands both biological and geometric understanding. This paper addresses the problem of constructing such 3D models for visualization. In this paper, we introduce a novel agent framework that mitigates the challenges of operating 3D modeling software by transforming user inputs, including natural language descriptions, research publication content, and textual descriptions of the existing objects and structures in the current scene, into modeling operations in a structured JSON format and final 3D results. The major technical contribution lies in the collaborative agent design that simultaneously supports model planning, execution, and novel user interaction design, such as interactive modeling execution and dynamic widget generation that fuse text and mouse interaction within the chat window. The framework further incorporates a customized modeling memory to enhance user interaction, featuring components such as personalized memory management, feedback collection, and skill library design. This modeling memory is leveraged to enable improved 3D modeling performance over time. The quantitative evaluation on our collected dataset showcases the effectiveness of our framework. We also develop a prototype tool, Chat Modeling, and demonstrate its usage through two modeling case studies. Our user study and expert interviews highlight the potential of our approach for use in scientific workflows.
\end{abstract}

\begin{IEEEkeywords}
Mesoscale modeling, molecular visualization, large language models
\end{IEEEkeywords}
\section{Introduction}
\begin{figure*}[t]
    \centering
  \includegraphics[width=0.85\linewidth]{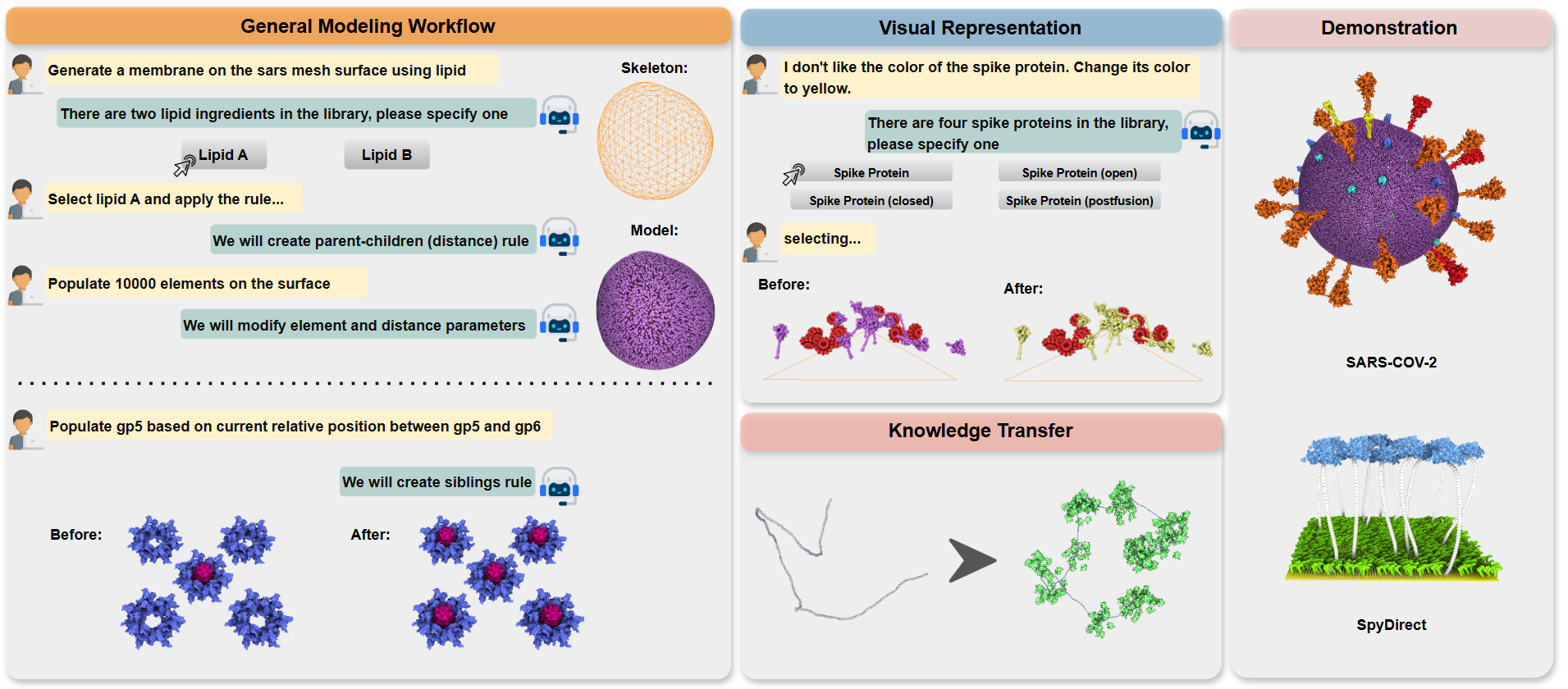}
   \caption{
   	Chat Modeling is an interactive agent framework that takes users' multimodal input and completes modeling operations. It can complete the general modeling workflow (left), modify visual representations, or transfer previous modeling templates to a new model (middle). We demonstrate our prototype through two complex biological structures modeled by different modeling modes (right).
   }
  
  \label{fig:teaser}
\end{figure*}
Modeling and visualization of the cellular mesoscale, the scale bridging atoms and molecules with whole cells, is necessarily an integrative process, since no existing experimental method can directly observe cellular mesoscale structures~\cite{goodsell2020art}. In practice, biologists construct mesoscale models by synthesizing literature evidence and composing known biological components into faithful and visualizable 3D representations. However, transforming such evidence into manipulable 3D mesoscale models remains labor-intensive, requiring both biological expertise and proficiency with complex modeling tools.
Current modeling tools, such as Maya~\cite{AutodeskMaya2023} and Blender~\cite{BlenderFoundation2023}, offer advanced modeling capabilities. However, these tools require users to interact with complex graphical user interfaces (GUIs) and possess a deep understanding of geometry and spatial relationships. This steep learning curve impedes users from effectively utilizing these tools, especially when modeling complex structures. For instance, structural biologists often struggle to translate their biological knowledge into the geometric and spatial constructs needed to create accurate 3D models~\cite{https://doi.org/10.1111/cgf.13072}. We target the paper-to-model gap by enabling scientists to turn structures described in their papers into faithful, visualizable 3D models with minimal effort. Our agent framework takes text commands, scientific publications, and the current scene context as input, then plans and executes the modeling steps through a natural-language interface, thereby obviating the need for manual specification of complex geometric relationships.

Large language models (LLMs), trained on vast amounts of textual data, have demonstrated exceptional capabilities in understanding and executing user instructions. Researchers have been leveraging LLMs in a variety of innovative contexts, including image reasoning and editing~\cite{suris2023vipergpt,gupta2023visual, michel2024object}, robotic manipulation~\cite{huang2023voxposer,wang2023prompt}, mesh and scene generation~\cite{Mohammad_Khalid_2022,sun20233d, Ma_2023_ICCV,feng2024layoutgpt}. These works utilized manually crafted prompts and well-designed multi-agent collaboration architectures, achieving strong performance, and current automatic prompt generation methods~\cite{zhou2022large} still cannot be applied to 3D scene scenarios. Despite their potential, three main challenges hinder the application of LLMs to modeling tasks. First, the unstructured textual outputs generated by LLMs cannot be directly parsed or processed by machines. To address this, we propose a novel JSON-based representation for modeling actions, which allows seamless parsing, interpretation, and execution by downstream systems. Second, traditional interactive modeling pipelines heavily rely on numerous object placement operations, which are difficult to express precisely through natural language. However, textual interfaces excel at specifying high-level modeling logic, such as defining procedural rules and adjusting their parameters. This observation naturally aligns LLM-driven modeling with a procedural modeling paradigm, enabling more rapid and flexible population of elements. Third, relying solely on text-based interactions between users and LLMs introduces inefficiencies in the modeling workflow, particularly when users need to perform frequent, fine-grained manipulations. To overcome this, we propose designing enhanced interaction mechanisms and a modeling memory to refine the modeling experience and improve modeling efficiency.

In collaboration with structural biologists, we conduct a formative study to gain insights into their typical modeling workflows and their specific needs for a natural language interface tailored to 3D modeling. Based on these insights, we propose Chat Modeling, a procedural modeling agent framework. This framework uses a set of specialized agents to interpret user inputs and carry out specific 3D biological structure modeling tasks, producing both a structured JSON recipe that controls the 3D scene and the corresponding updated 3D models. Each specialized agent focuses on one task and collaborates with the others. To better leverage user data, the framework incorporates a modeling memory designed to improve output accuracy, minimize redundant operations, and make effective use of previous user actions. To further enhance user interaction, we introduce several approaches within the modeling process.

We evaluate our framework through two approaches. First, we construct a benchmark dataset comprising 200 prompts to evaluate the framework’s ability to accurately interpret user commands into modeling operations. The results show that Chat Modeling outperforms general-purpose large language models in this task. Second, we develop a proof-of-concept prototype and conduct both exploratory user study and expert interview to collect qualitative feedback. These evaluations demonstrate the potential of Chat Modeling in supporting biological modeling tasks, underscoring its effectiveness in streamlining the modeling process and allowing users to focus on the creative and analytical aspects of their work. 

Our contributions can be summarized as follows: 
\begin{itemize}
    \item We present a conversational agent framework for mesoscale biological modeling that uses scientific literature to extract modeling parameters and supports interactive, step-by-step construction of biological structures.
    \item We design a new modeling recipe for procedural modeling, together with a corresponding interpreter that transforms users' high-level and abstract instructions into atomic modeling actions within the modeling software.
    \item We incorporate task-specific support mechanisms for biological modeling, including modeling memory, plan-stage interaction, and intent-conditioned widgets, to reduce user burden beyond text-only natural language interaction.
\end{itemize}
\section{Related Work}
\subsection{Natural Language Interfaces for Visualization}

Natural language interfaces have been widely studied as a way to lower the operational and learning barriers of visualization and modeling tools. Existing work spans a range of tasks, from querying structured data and editing visualizations to generating executable programs for visual reasoning, 3D scene authoring, and scientific visualization. A first line of work focuses on natural language interfaces for structured data visualization and visual analytics. Narechania \etal~\cite{narechania2020nl4dv} proposed NL4DV, which takes tabular data and natural language queries as input and returns an analytic specification in JSON format.  Mitra \etal~\cite{mitra2022facilitating} extended NL4DV to support multi-round conversational interaction through a conversation manager and query resolver. Wang \etal~\cite{wang2022towards} further studied natural language interfaces for visualization authoring, mapping user editing intents to concrete visualization editing actions through an interpreter that combines data-entry abstraction with a deep learning model. More recent systems use large language models to translate high-level language into executable scripts or modular programs for visualization and visual reasoning. Maddigan \etal~\cite{maddigan2023chat2vis} proposed Chat2VIS, which converts user queries and tabular data into Python scripts for chart generation. Gupta \etal~\cite{gupta2023visual} proposed VISPROG, where an LLM generates Python-like modular programs to invoke existing computer vision models for complex visual tasks. A more closely related direction connects natural language to 3D scene generation, mixed reality authoring, and scientific visualization. Jia \etal~\cite{11037292} proposed VOICE for scientific communication, which processes speech input to answer biological questions and generate visual explanations of biological structures. Sun \etal~\cite{sun20233d} introduced 3D-GPT, which uses LLMs to generate Python-like code for invoking functions from Infinigen~\cite{raistrick2023infinite} for 3D scene generation. De La Torre \etal~\cite{de2024llmr} proposed LLMR for LLM-assisted scene creation and modification in interactive mixed reality experiences. While these works move closer to spatial and 3D authoring, they primarily focus on scene composition, mixed reality content, or scientific communication. In contrast, biological modeling requires composing known biological components under literature-grounded, geometric, and quantitative constraints. Our work therefore focuses on a conversational agent framework that connects natural language, literature-derived evidence, and executable modeling operations for biological structures.

\subsection{Procedural Modeling}
Procedural modeling is an automation technique that generates complex 3D geometries and scenes through algorithms and rules rather than direct manipulation. Since its inception, procedural modeling has been extensively applied in fields such as computer graphics, game development, architectural design, and scientific visualization. The core advantage of this method lies in its efficiency in creating detailed and complex models, particularly suited for scenarios that require repetitive elements and structures, such as plants~\cite{prusinkiewicz2012algorithmic,guo2020inverse,niese2022procedural}, cloudscapes~\cite{webanck2018procedural}, roads~\cite{galin2010procedural}, street networks~\cite{parish2001procedural} and buildings ~\cite{muller2006procedural,schwarz2015advanced}.

Raistrick \etal~\cite{raistrick2023infinite} introduced Infinigen, a procedural system capable of generating photo-realistic 3D scenes of the natural world. This system can produce infinite shapes, materials, and scene compositions. Each asset, ranging from shapes to textures, is entirely procedural, created from scratch using randomized mathematical rules. These rules facilitate infinite variation and composition, generating highly diverse and unique scenes. Nguyen \etal~\cite{nguyen2020modeling} introduced a scientifically accurate procedural modeling method for creating 3D models of cellular mesoscale. Their method is implemented in MesoCraft, a desktop or web-based tool\footnote{\url{https://mesocraft.kaust.edu.sa}} for rapid modeling and visualization of sub-micron biological systems. Mesoscale biological modeling is particularly suited to procedural modeling due to the inherent repetitive elements and specific geometric relationships characterizing these structures. 

The steep learning curve of mastering complex geometric definitions in 3D procedural modeling, combined with intricate GUI operations, significantly hinders the user's modeling workflow. Therefore, this paper focuses on facilitating the modeling process by extracting explicit geometric definitions from users’ natural language inputs and implementing this workflow in the desktop version of MesoCraft.

\section{Formative Study} \label{sec: formative study}
Our primary objective is to develop a natural language interface to aid users in modeling intricate biological structures. This tool is designed to receive users’ inputs, assist them in formulating modeling tasks, and execute the corresponding modeling operations. It is specifically tailored for structural biologists who are new to specific modeling software and often encounter challenges while modeling complex biological structures. Modeling software typically incorporates complex graphical user interfaces and necessitates prior knowledge of spatial and geometric concepts. This combination leads to a steep and time-consuming learning curve for mastering modeling operations, underscoring the necessity for a tool that assists users by simplifying software interactions. A tool designed to streamline the modeling process would significantly benefit users. It would enable them to focus more on their work's creative and analytical aspects rather than on overcoming technical barriers.

To accurately identify the essential features our tool must incorporate, we have consulted with three experts (E1, E2, and E3), all of whom have prior experience in modeling complex biological structures. All experts possess backgrounds in biology, while E3 has additional computer science expertise. Notably, all three experts have experience modeling biological structures with MesoCraft, as demonstrated in previous works~\cite{nguyen2020modeling, NanovisG51:online, NanovisChloroplast:online}. We conducted individual interviews with each expert to understand the specific needs our tool should meet comprehensively.

Each interview was conducted over a span of approximately 90 minutes. Initially, the experts were asked to describe their previous modeling processes, enabling us to grasp their standard modeling workflows and the primary challenges they face. Following this, we delved into understanding their internal logic when determining rule types and asked the experts to construct example models in MesoCraft. The final segment of the interview focused on envisioning the ideal functionalities of a natural language interface designed to facilitate their modeling process. The experts were encouraged to specify the features they believe such a tool should offer and provide examples of how they would articulate these commands in natural language. We report below the design requirements from the interviews. 
\subsection{Design Requirements}
\textbf{DR1--Consistent with typical modeling pipelines}: \textit{The interface should align with experts’ typical modeling workflows.} Expert interviews indicate a consistent two-step process. First, users \emph{create rules} (e.g., ``Create a lipid bilayer membrane on the virus skeleton surface''). Given that biological ingredients often have similar names and users may prefer not to specify the full names, there is an expectation to intelligently narrow down the options and present a selection of choices for the user to choose from. Second, users \emph{adjust rule parameters} in several iterations, such as `I would like to populate 1000 lipids on the skeleton surface,’ explicitly setting the count while implicitly setting the lipid–skeleton gap to~0.  Separating rule creation from tuning reflects real practice, where frequent parameter tweaks are common.  

\textbf{DR2--Iterative rule refinement}: \textit{Users can reapply any rule at any time to tweak its parameters and freely reorder its execution within the rule list.} Biological models usually combine many rules, and their execution order is critical: ingredients occupying key positions must be placed first to avoid collisions.  Hence, the interface must let users reorder or recall rules at will, enabling precise, incremental refinement of the structure.

\textbf{DR3--Long-term refinement \& reusability}: \textit{The interface should be adaptive to user habits and reuse similar operations.} After providing the pairs of natural language inputs and modeling rules, experts indicated that this collected data might not encompass the full spectrum of modeling scenarios or the variety of language expressions used. Structural biologists may lack a comprehensive understanding of geometric positional relationships or possess their own unique terminology for describing these rules. They suggested an enhanced approach: collecting specific user cases and directly incorporating an approval or correction mechanism into the tool. Also, experts further emphasized that, unlike general 3D modeling, which usually creates novel objects tailored to diverse user requirements, biomolecular modeling is driven by experimentally determined or literature-sourced structural knowledge and focuses on faithfully reconstructing known biological assemblies. Because many biomolecules share recurring structural motifs, the ability to efficiently reuse previously constructed models is especially critical.

\textbf{DR4--Multi-stage modeling workflow}: 
\textit{The interface should support a clearly structured, multi-stage modeling workflow that makes high-level biological modeling tasks executable and controllable.} Rather than executing a single user-issued command, the system should decompose user-derived intentions into an explicit sequence of modeling operations. While the idea of deriving planning from high-level descriptions has been explored in prior natural-language-driven systems, this requirement centers on adapting such capabilities to the particular demands of biological modeling, such as extracting modeling evidence from publications, generating a high-level modeling plan, exposing intermediate plans for user inspection, and supporting validation and refinement of the resulting model. In our context, the interface should surface these otherwise implicit steps as revisable stages in a structured workflow. This organization helps reduce the need for users to specify low-level geometric details and makes it easier to manage and refine complex procedural modeling pipelines.

\textbf{DR5--Multimodal interaction workflow}: \textit{The interface must go beyond text-only interaction, harnessing the complementary strengths of mouse and text input.} Mouse actions excel at coarse spatial manipulations-such as roughly rotating or translating a structure- whereas text commands provide precise control (e.g., ``rotate 30° to the right’’) and allow users to issue high-level instructions that bundle several elementary operations. By dynamically selecting the modality that minimizes user effort for each task, the system streamlines the modeling workflow and aligns with users’ natural working habits.

\textbf{DR6--Multimodal input}: \textit{The interface should support multimodal input tailored to biological modeling workflows.} In conventional 3D scene modeling tasks, users typically construct scenes based on their imagination, focusing on the shapes of specific objects and performing multiple rounds of adjustments. Sketches and text are often combined to describe the desired object shapes. In contrast, 3D biological modeling aims to reconstruct known biological structures using modeling tools. In this context, the shapes and approximate locations of proteins (e.g., within or on the membrane) are already determined. Users are more concerned with information such as the quantity of proteins and their spatial distributions, which are often obtained from experiments. Therefore, experts suggest that, beyond traditional text input, the agent is better to support extracting critical data, such as protein counts, from publications like scientific papers, enabling richer and more versatile multimodal input.

To make the relationship between the design requirements and our contributions explicit, we summarize how each requirement informs the proposed system design and corresponding contributions. DR1 and DR2 motivate the design of our procedural modeling recipe and interpreter, which separate rule creation from parameter tuning and allow rules to be reapplied and reordered during modeling. DR3 directly motivates the customized modeling memory, which captures user preferences, collects feedback, and enables reuse of prior modeling operations in literature-grounded biological modeling tasks. DR4 motivates the overall conversational agent framework and the explicit decomposition of high-level biological intentions into a structured, inspectable, and revisable plan. DR5 and DR6 motivate our interaction designs during plan execution, including intent-conditioned widgets and the integration of scientific literature as an input modality.
\section{Method Overview}\label{sec:overview}

\begin{algorithm}[t]
\caption{Chat Modeling}\label{alg:chat_modeling}
\KwInput{	$\mathbf{u}$: user input; 
\quad $\Omega$: current scene raw;
$\mathbf{m}$: modeling memory;}
\KwRequire{$
  \begin{aligned}[t]        
    &S(s\mid \Omega)                     :\ \text{Scene Summarizer}            \\[2pt]
    &P(u_1,\ldots,u_N\mid u,s,m)             :\ \text{Modeling Planner}          \\[2pt]
    &G( j \mid u,e,m)      :\ \text{Recipe Generation Agent}     \\[2pt]
    &C( e \mid j)         :\ \text{Recipe Correction} \\[2pt]
    &I( \mathcal{R},\mathcal{P},\mathcal{A} \mid j,m):\ \text{Recipe Interpreter}
  \end{aligned}$}
$s \sim S(\cdot \mid\Omega)$ 

$(u_1,\ldots,u_N) \sim P(\cdot \mid u,s,m)$ \hfill \textit{//generate plan from multi-turn dialogue}

\For{$i \gets 1$ \KwTo $N$}
{
 $j \sim G(\cdot \mid u_i,m)$ \hfill \textit{//generate recipe based on the step}
 
 $e\sim C( \cdot \mid j)$ 
 
\While{$e$}{
        $j \sim G(\cdot \mid u_i,e,m)$ \hfill \textit{//correct json by the error info} 
        
        $e\sim C( \cdot \mid j)$ \hfill \textit{//check the error again}
  }
$\mathcal{R}_i,\mathcal{P}_i,\mathcal{A}_i \sim I(\cdot \mid j,m)$ \hfill \textit{//Interpret the recipe to operations}

$\mathcal{M} \leftarrow \text{RunInMesoCraft}(\mathcal{R}_i,\mathcal{P}_i,\mathcal{A}_i)$
}

\BlankLine
\textbf{Notation:}
$\mathbf{u_i}$ denotes the $i$-th step in the modeling plan;\\
$j$ is a structured recipe representation;\\
$e$ indicates validation errors;\\
$\mathcal{R}, \mathcal{P}, \mathcal{A}$ denote modeling rules, parameters, and actions, respectively.
\end{algorithm}

Recent advances in LLMs have greatly improved natural language understanding and instruction following capabilities. Leveraging these capabilities, we aim to propose an agent framework that closes the gap between free-form textual commands and the mouse-driven workflows of existing modeling software. Informed by DR4 (multi-stage modeling workflow) from our formative study, the framework not only assists users in devising a step-by-step modeling plan but also translates each planned step, or raw user instructions, directly into concrete modeling operations.

An effective strategy to bridge this gap involves adopting the procedural modeling paradigm and creating an intermediary component layer. This layer serve as a conduit between users and the modeling software. Essentially, it would translate human-readable instructions into a software-processable recipe, acting as a translator between the user's textual descriptions and the software's command syntax. Here, a recipe denotes a structured model specification-listing the biological components, their quantities, and relevant constraints/parameters-that the modeling system can directly execute.

\begin{figure}[t]
    \centering
    \includegraphics[width=0.7\linewidth]{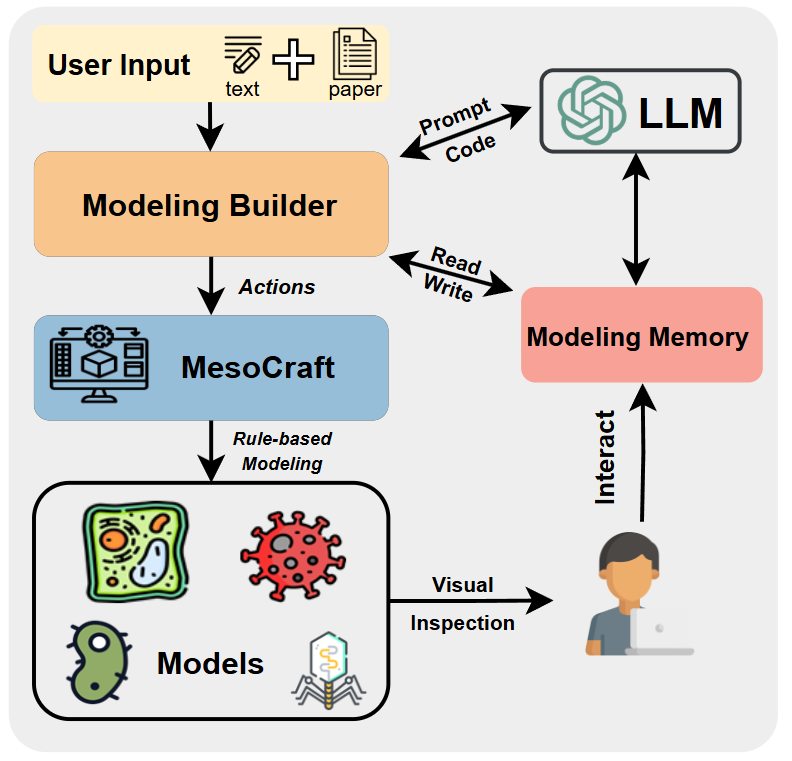}
    \caption{
    The framework starts with user input, processed by the Modeling Builder to generate modeling plans. MesoCraft then models biological structures from these plans. Users visually inspect results and interact with the modeling memory to improve the whole framework.
    }
    \label{fig: overview}
\end{figure}

The overview of our framework is illustrated in \autoref{fig: overview} and formatted in \autoref{alg:chat_modeling}. Our workflow starts with a user’s input $u$. To fulfill DR6 (multimodal support) that biological modeling always aims to reconstruct known biological structures, we incorporate two types of input: textual descriptions and scientific publications such as research papers. User input together with the scene information $s$, is sent to the modeling builder, the bridging layer between the user and the MesoCraft. We define $s$ as the current scene state, which consists of a text-based description of the objects, and structures already present in the scene. The modeling builder performs two complementary tasks: (1) through multi-turn dialogue, it distills the user’s description into an ordered, high-level modeling plan; and (2) it converts each plan step $u_i$ into machine-readable JSON recipe $j$ that is subsequently translated into native MesoCraft operations, such as rules $\mathcal{R}_i$, parameters $\mathcal{P}_i$ and modeling actions $\mathcal{A}_i$. MesoCraft then executes these operations, generating the requested biological structures $\mathcal{M}$ for the user’s inspection and refinement. To enable continual improvement, the framework also features a dedicated modeling memory $m$ that aggregates data from both the user’s visual feedback and the dialogue history to enhance planning and execution quality across future modeling sessions.

\begin{figure}[ht]
    \centering
    \includegraphics[width=\linewidth]{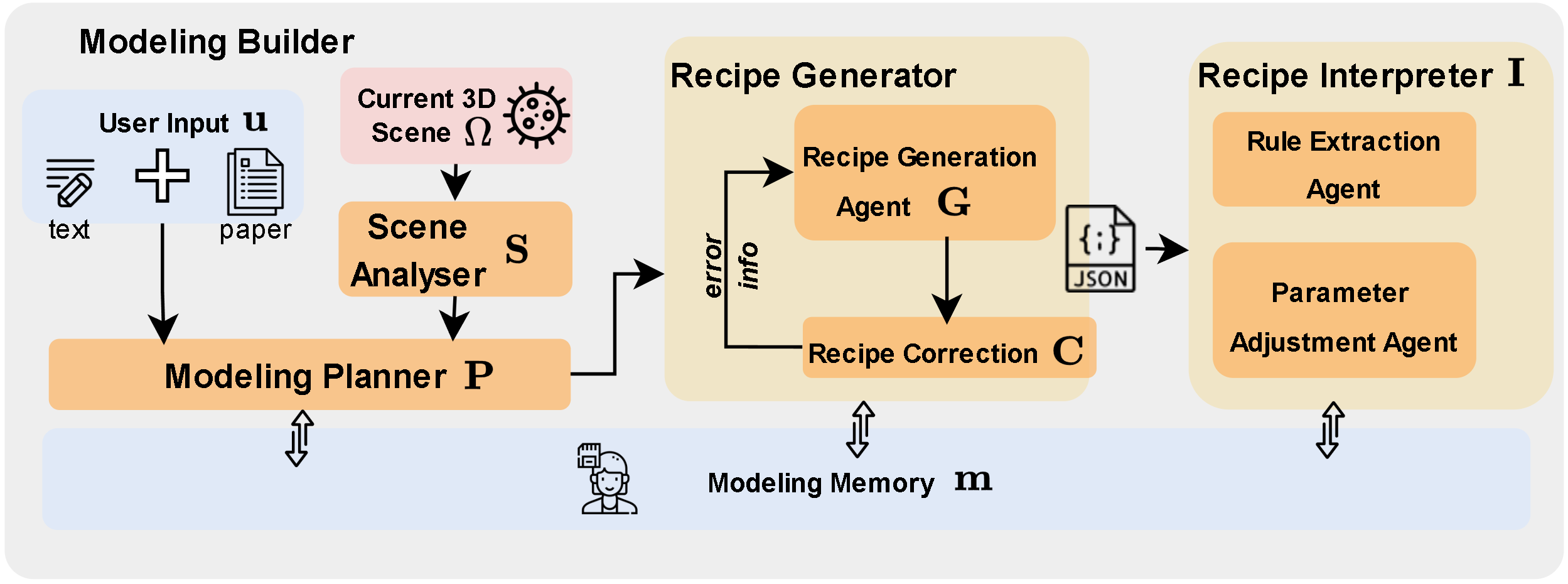}
    \caption{The modeling builder consists of the modeling planner, recipe generator, and recipe interpreter. The recipe generator creates validated recipe based on the plans from the modeling planner, while the recipe interpreter converts this recipe into procedural modeling actions.}
    \label{fig:Modeling Builder}
\end{figure}

The modeling builder, illustrated in \autoref{fig:Modeling Builder}, consists of three principal components, \ie, the modeling planner $P$, the recipe generator and the recipe interpreter $I$. The modeling planner takes user input $u$ and scene description $s$ as input and generates a modeling plan $U=\{u_{1},\ldots, u_{N}\}$,  where each $u_i$ represents an individual modeling step. The recipe generator transforms steps into validated and executable recipe snippets $j$. Following this, the recipe interpreter $I$ parses and interprets these validated scripts. It is crucial in translating abstract recipe into concrete operations within MesoCraft, effectively bridging the gap between user intent and software execution.

The recipe generator component includes two key operations: recipe generation agent $G$ and recipe correction operation $C$. Utilizing LLMs, it deciphers user commands to produce accurate recipe snippets encapsulated in a novel JSON format, tailored for complex biological modeling workflows. After recipe generation, the recipe correction process ensures the accuracy and effectiveness of these snippets by feeding any detected errors back into the LLMs for modification. The recipe interpreter involves three main components: a rule extraction agent, which identifies rule types $\mathcal{R}_i$ based on the geometrical and spatial relationships in the step; parameter adjustment agent, which allows for fine-tuning these rule parameters $\mathcal{P}_i$; and functionalities for visual representation modifications and other essential actions $\mathcal{A}_i$ within the general modeling workflow. Once interpreted by the recipe interpreter $I$, the actions are executed within MesoCraft to create complex biological models, which are then presented back to users.

As mandated by DR3 (long‑term refinement and reusability), structural biologists articulate geometric and spatial relationships in highly diverse ways, which can reduce the reliability of the framework’s natural-language-driven modeling operations. To accommodate this variability, we embed a modeling memory into our workflow. The system persistently logs users’ interactions within the modeling builder and records their visual-inspection feedback on generated results. These traces are subsequently leveraged to iteratively refine future outputs and to recall previously validated operations, thereby lightening users’ modeling workload and enhancing overall accuracy.

To address the issue of text-only interactions being suboptimal for certain scenarios, as outlined by DR5 (multimodal interaction workflow), we incorporate multiple interaction strategies during the plan execution stage. This hybrid approach enhances the modeling process, reducing user effort by enabling more intuitive and efficient interactions.

\section{Modeling Builder}\label{sec:modeling translator} 
This section details the implementation of each component within the modeling builder, which generates the modeling plan and translates natural language steps into executable actions of the modeling software. 

\subsection{Memory-enhanced Few-shot Prompting}
\begin{figure}[t]
    \centering
    \includegraphics[width=0.85\linewidth]{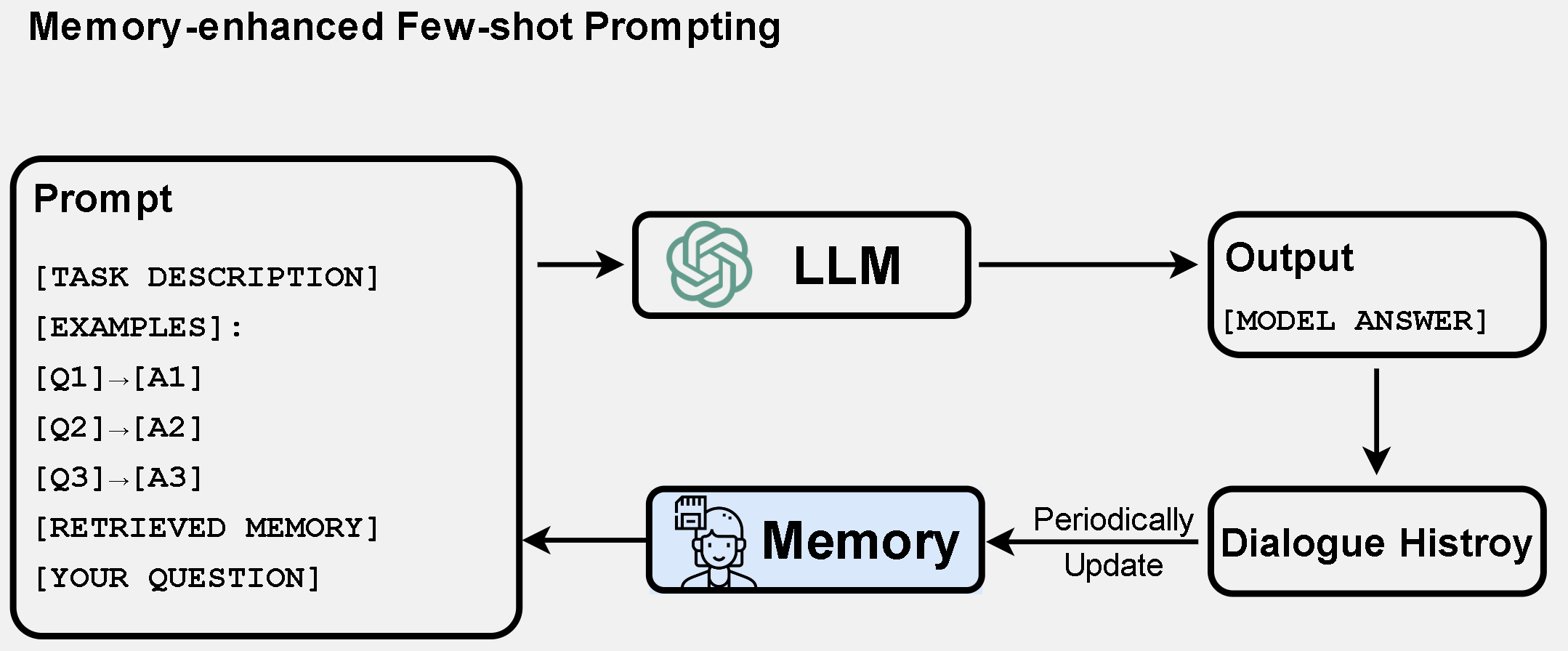}
     \caption{Memory-enhanced few-shot prompting setting. The prompt includes a task description, initial examples, and retrieved memory. Memory is periodically updated from the dialogue history.}
    \label{fig:prompt}
\end{figure}

As outlined in \autoref{sec:overview}, our approach leverages LLM-based agents to handle various tasks in the modeling builder. While LLMs excel at grasping general semantics, their application in task-specific scenarios requires good prompts. Prompting is a technique that involves the careful crafting of inputs to the model to enhance task performance. Importantly, this process does not involve modifying the model's weights, ensuring that the LLM can be effectively utilized for specific tasks through strategic input manipulation alone. Our methodology adopts a memory-enhanced few-shot prompting setting, demonstrated in \autoref{fig:prompt}. The prompt contains a concise task description and a few task-specific examples. Previous work~\cite{brown2020language} verified that a few examples could significantly improve LLM performance. We collected these examples from experts in the formative study. 

To accommodate the diverse ways in which users articulate modeling intent, and to enable our framework to improve its accuracy over time, we append a user-specific, dynamic memory block to the prompts. This block either supplements the task-specific examples discussed above or encodes habitual modeling preferences unique to the current user. As the modeling memory is updated and queried, the contents of this block are updated periodically, allowing each prompt to be continuously tailored to the user’s needs. The design of the memory component is described in detail in \autoref{sec:memory}, and complete prompt templates of different agents are provided in the supplementary material.

\subsection{Modeling Planning}
The modeling builder initiates with a planning stage that integrates the user's multi-turn dialogue with the current on-screen context to generate a detailed modeling plan. This plan serves as the blueprint for downstream agents to construct the target biological structure. The planning stage involves two cooperative agents: the scene summarizer agent, which converts raw screen data into a textual scene description, and the modeling planner agent, which fuses the user's dialogue with the scene description to produce the finalized modeling plan.
\subsubsection{Scene Summarizer}
Our modeling planning stage operates on the current scene data, which stores heterogeneous information such as the name, category, and spatial pose of every instance. Because these attributes are scattered throughout complex scene data, we introduce a scene summarizer agent~$S$ that converts the raw scene data~$\Omega$ into a concise textual summary of scene information~$s$ for the planner. Concretely, the raw scene data~$\Omega$ is extracted from the MesoCraft scene as a JSON object that lists essential information such as each instance’s name, type, position, rotation, together with the global skeleton and per-instance hierarchy metadata. When a biological structure contains numerous copies of the same protein, the raw scene data $\Omega$ omits per-instance pose details (e.g., exact positions and orientations) to keep the summary compact. One example of the input and output of the scene summarizer agent is shown in \autoref{fig:scenesummrizer}. The detailed agent prompt is provided in supplementary section 2, which includes the task description and several illustrative examples. 
\begin{figure}[h]
    \centering
\includegraphics[width=1\linewidth]{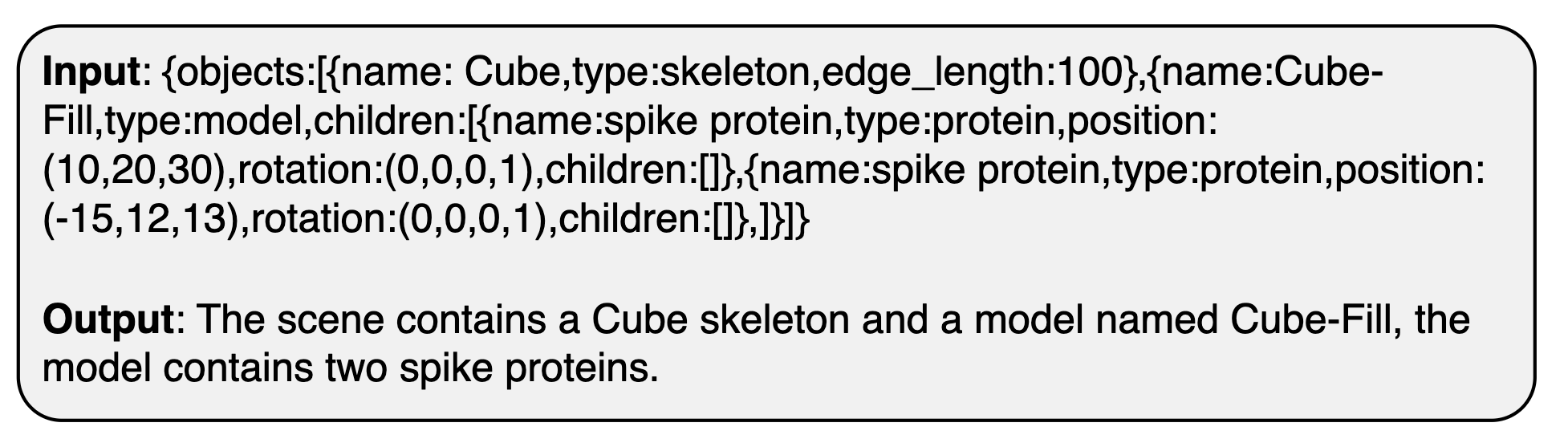}
    \caption{An example illustrating the input and output of the Scene Summarizer, which converts raw scene data into semantic scene information.}
    \label{fig:scenesummrizer}
\end{figure}

\subsubsection{Modeling Planner}
In traditional mouse-driven GUI-based workflows, users need to first conceptualize the model’s global structure, shape, and the spatial relations among its constituent instances. Powered by the broad biological knowledge embedded in LLMs, we introduce a modeling planner agent $P$ to fulfill DR4 (multi-stage modeling workflow). Through a multi-turn dialogue, the planner, grounded in the current scene context $s$, returns an initial modeling proposal. The user may then iteratively revise this proposal until a finalized plan is obtained. 

\begin{equation}
  P(u_1,\dots,u_N \mid u,s,m)=
  \prod_{i=1}^N P\!\bigl(u_i \mid u,s,m,u_1,\dots,u_{i-1}\bigr),
\end{equation}

where $u$ denotes the user’s query, $s$ the textual scene summary, and $m$ the memory. The variable $u_i$ represents the $i^{\text{th}}$ step proposed from the plan, and the planner’s output is the sequence $U=\{u_1,\dots,u_N\}$. As specified by DR6 (multimodal input), 3D biological modeling focuses on reconstructing known biological structures, with particular emphasis on the quantity of proteins and their spatial distributions, which are of primary concern to users. Consequently, the user query $u$ consists of two components: textual descriptions provided by the user, and scientific publications such as research papers. The scientific publications, which are optional, are primarily used to supply accurate parameters. In their absence, the parameters are inferred either from the LLM's internal knowledge or directly from user input. To personalize planning, the memory $m$ encodes user-specific context such as modeling preferences, enabling the planner to adapt its suggestions to the current user. When scientific publications are unavailable or insufficient, parameters may be inferred from the LLM’s internal knowledge, which introduces the risk of hallucination. To mitigate this issue, all parameters inferred from literature are explicitly marked in the dialogue interface, making their provenance transparent to the user. Importantly, these inferred parameters are treated as tentative initial conditions rather than fixed constraints: they only initialize the planning process and can be inspected, modified, or overridden by the user through subsequent conversations. This design enables efficient human verification and correction, ensuring that potential hallucinations do not propagate into the final modeling results.

The detailed modeling planner agent prompt is provided in supplementary section 1, which includes the task description, detailed guidelines, and the basic logic of the MesoCraft modeling process, current scene information, user modeling preference, and a few modeling plan examples. The current scene information is provided by the scene summarizer agent, and the user modeling preference is from the modeling memory.

\subsection{Recipe Generator}
The recipe generator comprises two key operations, \ie, recipe generation agent and recipe correction operation. Through iterative adoption of these operations, the recipe generator forwards validated recipes to the recipe interpreter for further processing.
\subsubsection{Recipe Generation}
The recipe generation operations extract user intents into machine-readable recipes, aiming to precisely mirror the user's modeling workflow actions and accommodate the requirements for visual representation modifications. The chosen recipe format must be easily parsed and executable for subsequent interpretation. Based on these criteria, we propose a new JSON format covering user intents collected from the formative study. JSON is a standardized yet flexible format, ideally suited to handle the complexities of modeling tasks. The specifics of the recipe format are illustrated in \autoref{code generation}.

Our format incorporates multiple intents to support the general modeling workflow within MesoCraft comprehensively, as requested by DR1 (consistent with typical modeling pipelines). Each modeling action is predicated upon creating and applying specific rules. These rules comprise one or several ingredients and an optional skeleton structure to guide the geometrical arrangement of populated instances. We introduce two pivotal intents, \ie, \textit{selectIngredient} and \textit{selectSkeleton}, to support modeling basic operations. Because a single instruction may reference multiple items, both accept arrays. We also add \textit{createRule} and \textit{editRule} for creating new rules and modifying parameters. Given the difficulty of inferring rule types and numerous tunables, the recipe interpreter includes a rule extraction agent and a parameter adjustment agent. The recipe retains rule and parameter descriptions for downstream processing. Finally, \textit{saveModel} and \textit{loadModel} are used for model save and load. Natural language conveys precision that mouse clicks cannot, for example, positions by amino acid index in biology. To leverage this, we add \textit{updatePivot} and \textit{updatePosition}: users can set an instance’s pivot by amino acid index and update its position by specifying spatial relations between amino acids across instances. For visual representation edits (informed by our formative study), we include \textit{highlightIngredient}, \textit{modifyColor}, and \textit{changeMode}. \textit{highlightIngredient} emphasizes selected ingredients; \textit{modifyColor} alters their colors; both accept arrays to batch multiple changes per instruction. \textit{changeMode} switches MesoCraft rendering among protein, chain, and atomistic-level views, each offering a distinct perspective for analysis or presentation.
\begin{lstlisting}[language=json,label=code generation,caption=Syntax format of the recipe generation agent.]
{"selectIngredient":[
    {"ingredient":(Str)ingredient_name},],
"selectSkeleton":[
    {"skeleton":(Str)skeleton_name},],
"createRule":(Str)descriptions_to_the_rule,
"editRule":(Str)descriptions_to_edit_the_rule_parameters,
"saveModel":(Bool)if_save_model,
"loadModel": (Bool)if_load_model,
"updatePivot":{
      "chainId":(Int)chain_index,
      "residueId":(Int)residue/amino_acid_index,},
"updatePosition": [
    {"mainIngredient":(Str)ingredient_name ,
       "chainId":(Int)chain_index,
       "residueId":(Int)residue/amino_acid_index},
    {"subIngredient":(Str)ingredient_name,
     "chainId":(Int)chain_index,
     "residueId":(Int)residue/amino_acid_index}],
"highlightIngredient": [
    {"ingredient":(Str)ingredient_name }],
"modifyColor": [
    {"ingredient":(Str)ingredient_name,
     "color":(vector3D)ingredient_color}],
"changeMode":(Str)rendering_mode_name_to_change,
"labeling":(Bool)if_labeling_enabled}
\end{lstlisting}

The detailed recipe generation agent prompt is provided in supplementary section 3, which includes the entire task description, detailed JSON format, reasoning logic about the recipe generation, and a few recipe generation examples together with the user feedback from the memory system.

\subsubsection{Recipe Correction}

\begin{figure}[t]
    \centering
    \includegraphics[width=0.85\linewidth]{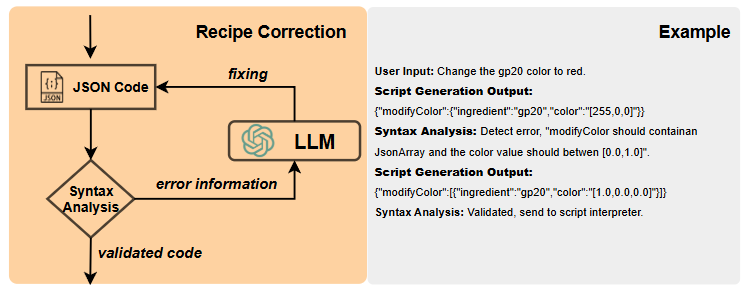}
    \caption{The recipe correction operation involves an iterative process for syntax analysis and error fixing.}
    \label{fig:code correction}
\end{figure}

We incorporate a recipe correction stage into the generator to handle small but consequential errors in LLM outputs. Although recent LLMs perform well at recipe generation~\cite{chen2021evaluating}, domain-specific constraints, such as our custom schema for complex biological models, still lead to occasional format discrepancies. As shown in \autoref{fig:code correction}, we implement a syntax analysis module in MesoCraft, containing both heuristic rules and a JSON schema linter to check the recipe code. This module functions as a quality-assurance gate that enforces strict conformance before any recipe reaches the interpreter: it reviews each recipe for parseability (syntactic validity), validates array-object structure for intents such as \textit{selectIngredient} and \textit{modifyColor}, and verifies that all values match the predefined data types in our schema. When violations are detected, we attach concise, manually defined error descriptors and return the recipe to the generator for up to $n{=}2$ revision rounds; each round addresses only the reported issues and then re-runs validation. Once all checks pass, the corrected recipe proceeds to the interpreter. The example on the right of \autoref{fig:code correction} shows two syntax errors that are flagged, corrected in one pass, and subsequently validated.

\subsection{Recipe Interpreter}
Upon receiving the validated recipe, the recipe interpreter parses it and executes it into specific actions within MesoCraft. Actions like color modification and ingredient selection are directly executed using the key-value pairs from the scripts. Given the complexity of identifying rule types for rule creation and the necessary parameters for rule editing, we design separate agents, \ie, the rule extraction agent and the parameter adjustment agent.

\subsubsection{Rule Extraction Agent}
\begin{figure}[t]
    \centering
    \includegraphics[width=0.85\linewidth]{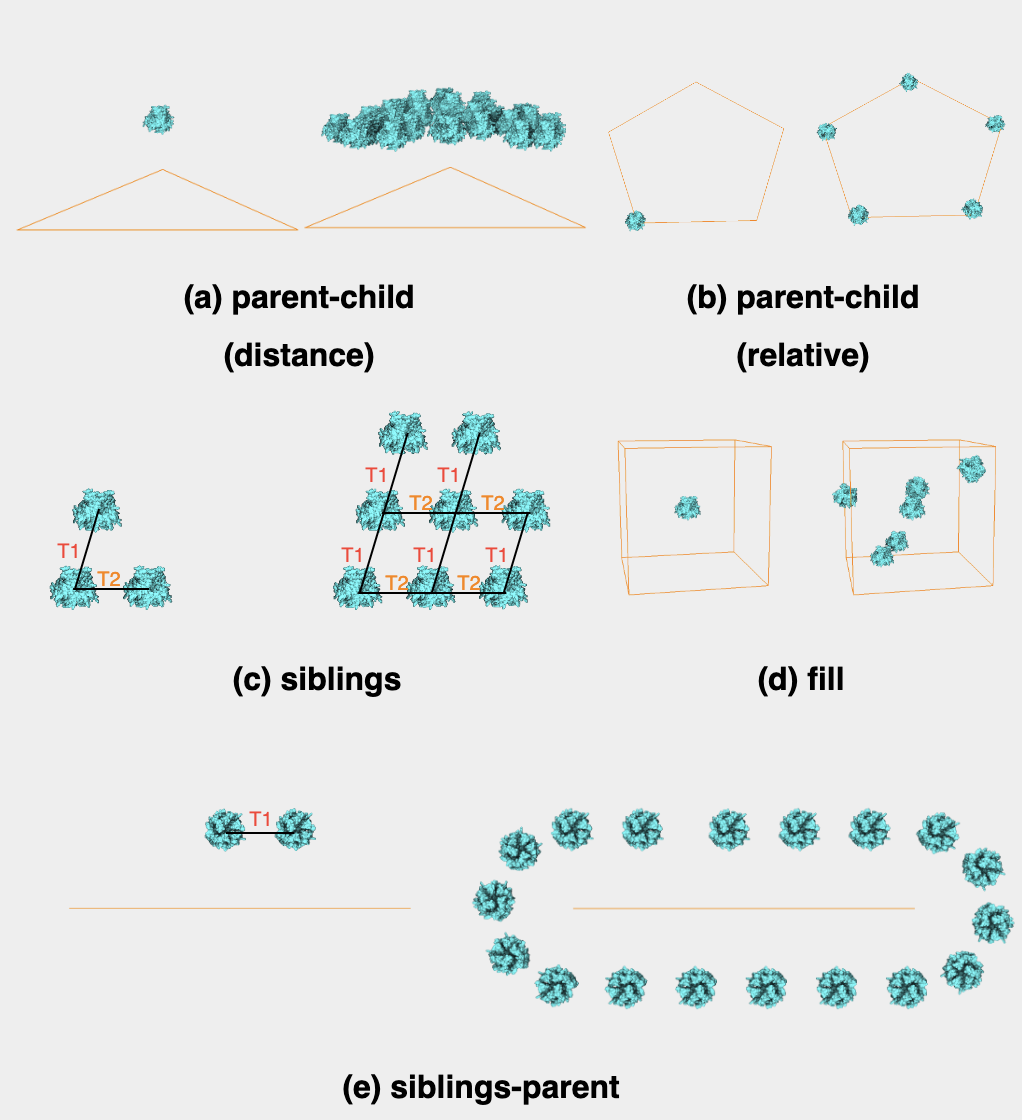}
    \caption{
    Examples of rule types: for each rule, the left image shows rule creation and the right image shows the outcome after rule application.
    }
    \label{fig: rule examples}
\end{figure}

Our framework operates on the rule system of MesoCraft, which represents procedural modeling operations through six classes of geometric rules: \textit{parent--child (distance)}, \textit{parent--child (relative)}, \textit{siblings}, \textit{siblings--parent}, \textit{fill}, and \textit{connection} rules. The rule extraction agent maps high-level user instructions or recipe descriptions to the appropriate rule type and its parameters, so that the modeling intent can be executed by MesoCraft. For example, a parent--child distance rule can place new elements at a specified distance from a skeleton surface, while a siblings rule can infer transformations between existing elements and reuse them to generate additional instances. Other rule types similarly encode geometric relations between elements, skeletons, or trajectories, enabling MesoCraft to populate and organize biological components in a controllable procedural manner. Representative examples are shown in \autoref{fig: rule examples}.

In MesoCraft, the first step in creating a rule is to determine its class. This choice depends on multiple factors, including the number of ingredients involved, the dimensionality of the skeleton (2D or 3D), and the spatial relationships between ingredients and the skeleton. Because this decision is non-trivial, we treat rule-type selection as a separate subtask and introduce a rule extraction agent within the recipe interpreter. More specifically, the agent’s prompt explicitly encodes the descriptions of each rule type together with the corresponding decision criteria, and is further augmented with system-provided examples and relevant rule-extraction-related user feedback from user memory. This prompt design guides the agent to determine the most appropriate rule class. Given a textual rule description, this agent predicts the appropriate rule class, whose parameters are then refined by the parameter adjustment agent. The detailed rule extraction agent prompt is listed in supplementary section 4. 
\subsubsection{Parameter Adjustment Agent} 
An essential operation of the recipe interpreter involves parameter adjustment. As highlighted in DR2 (iterative rule refinement), users often use multiple iterations to fine-tune rule parameters, ensuring the modeling output meets their expectations. The design of the parameter adjustment agent follows the recipe generation paradigm, introducing a novel format to encapsulate the various rule parameters, as depicted in \autoref{Parameter Adjustment}. This process begins with processing user instructions to generate a corresponding validated recipe via the recipe generator. Subsequently, the recipe interpreter parses and interprets this validated recipe, facilitating precise adjustments to the model. 

\begin{lstlisting}[language=json,label=Parameter Adjustment,caption=Syntax format of parameter adjustment operation.]
{
"elements":(Int)number_of_elements,
"distance":(Double)distance_to_the_skeleton,
"collisionDetection":(Bool)if_collision_detection_enabled,
"space":(Int)occupied_space,
"alignDirection":(Str)elements_aligned_direction,
"length":(Double)generated_curve_length,
"curve":(Str)generated_curve_type,
"tweaking":(Str)tweaking_directions,
"std":(Double)standard_deviation_for_the_distance
}
\end{lstlisting}

We encode a compact set of parameters in the recipe format: target rule elements, a collision-detection flag, mean/SD of instance–skeleton distance, alignment (along or opposite the skeleton normal), generated-curve length, and tweak angle. We also include high-level intent into the parameters. For example, a \textit{space} parameter specifies the fraction of the 3D skeleton to occupy; given the skeleton and ingredient volumes (approximating the latter with a bounding sphere), we compute the corresponding element counts and sizes. The detailed parameter adjustment agent prompt is listed in supplementary section 5, which contains a task description, the detailed format of the parameter recipe, and examples collected both from the formative study and user feedback from the modeling memory.

\section{Modeling Memory Design} \label{sec:memory}

The modeling builder supports multi-turn, dialogue-driven biological modeling. To support DR3 (long-term refinement \& reusability), we incorporate a lightweight modeling memory that helps the system adapt to user preferences, preserve corrective feedback, and reuse prior operations across sessions. Rather than introducing a new general-purpose memory architecture, our design adapts common memory mechanisms in conversational AI systems~\cite{lewis2021retrievalaugmentedgenerationknowledgeintensivenlp,huang2023memory,zhong2024memorybank} to the procedural workflow of biological modeling.

\begin{figure}[h]
    \centering
    \includegraphics[width=\linewidth]{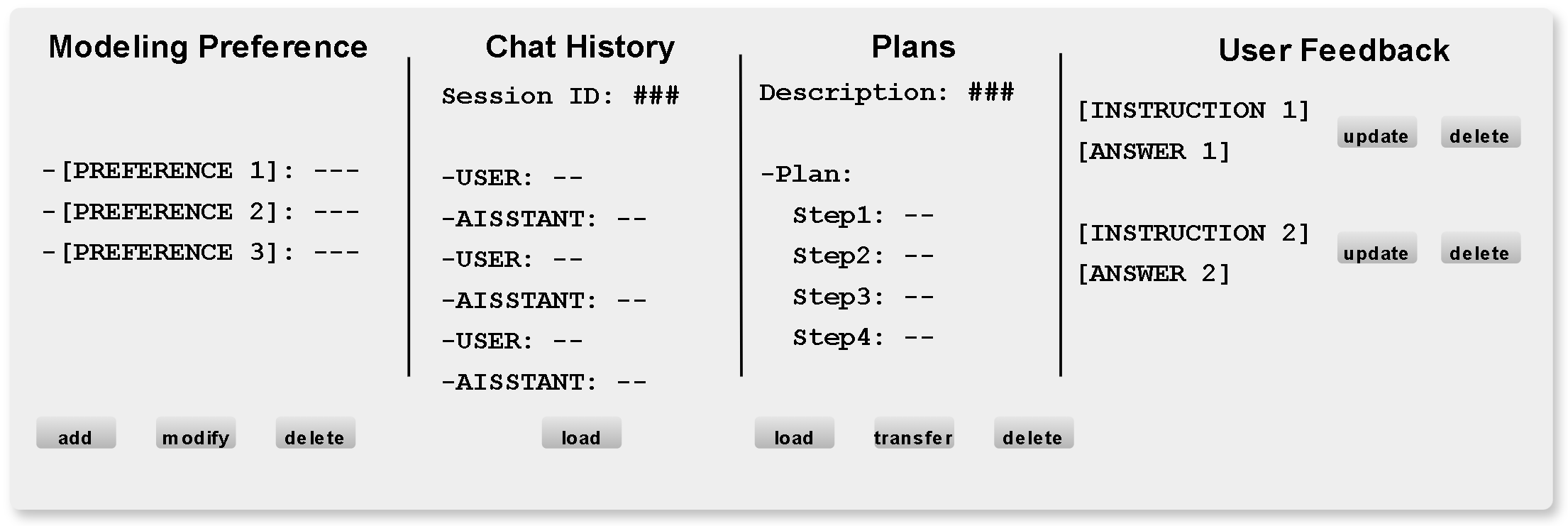}
    \caption{Demonstration of the memory management widget.}
    \label{fig:memory}
\end{figure}

As illustrated in \autoref{fig:memory}, the memory widget organizes four types of information. 
Modeling preferences store user-specific conventions, such as preferred parameter settings or naming styles. They are extracted from dialogue logs after multiple planning rounds and can be viewed, edited, or deleted by the user. User feedback records explicit corrections reported during plan execution, such as incorrect rule interpretation or parameter adjustment. The system converts such corrections into instruction--action pairs and routes them to the corresponding agent, including the recipe generation, rule extraction, or parameter adjustment agent. Previously executed plans store complete plan descriptions and executable plans as reusable templates for similar biological structures. Chat history maintains planner-relevant conversational context indexed by session IDs, enabling users to resume prior modeling sessions.

These memory types are consumed differently during modeling. Modeling preferences are appended to the planner prompt to condition planning and parameter selection. User feedback is injected into the corresponding agent prompt as few-shot instruction--action examples. Previously executed plans can be selected as templates to bootstrap new modeling processes, while chat history provides context for continuing prior sessions.

For example, if the recipe generation agent produces a valid JSON recipe but the rule type is misinterpreted by the rule extraction agent, the user can report the error and provide the correct interpretation. The correction is stored as feedback memory for the rule extraction agent and later reused as a few-shot example when similar rule descriptions appear.

\section{User Interaction Enhancements}
In the preceding sections, we introduce our agent framework grounded in the design of a modeling builder and a modeling memory. The framework is designed to accept text input and research papers as its primary inputs. However, as articulated by DR5 (Multimodal Interaction Workflow), relying solely on text to control the modeling process is often cumbersome. To better leverage the complementary strengths of natural-language specification and mouse-based direct manipulation, we employ a suite of multimodal interaction–augmentation strategies that more effectively support users in completing modeling tasks.

\textbf{Intent-conditioned widgets:} We design interaction modes by estimating the user effort associated with each operation and selecting the modality that minimizes this cost. For instance, continuous geometric adjustments (e.g., translation, rotation, scaling) are mapped to mouse-based direct manipulation. Discrete visual-representation edits (e.g., color) are offered through concise text commands or lightweight form controls, which reduce pointing and menu navigation; once a visual change is triggered via text, the user can continue refining it in a dialog using a color picker without issuing additional textual commands. In the biological domain, where protein names are often similar, we generate selection buttons in the dialog whenever user disambiguation is required, and execution only proceeds after the user makes a selection. To support the DR2 (iterative rule-refinement paradigm) in biological modeling, where each rule undergoes multiple rounds of fine-tuning to meet user expectations, we dynamically generate control buttons when a rule is applied. Users may then adjust rule parameters either through natural-language specification or via interactive widgets in the dialog (e.g., sliders, numeric inputs), thereby minimizing effort. Overall, intent-conditioned widgets allow users to perform modeling operations either purely through natural-language instructions in the dialog or by quickly reapplying the same operation via intent-conditioned widgets. This not only reduces user effort but also ensures that, after multiple iterations, users can conveniently reapply previously executed operations.

\begin{figure}[h]
    \centering
    \includegraphics[width=\linewidth]{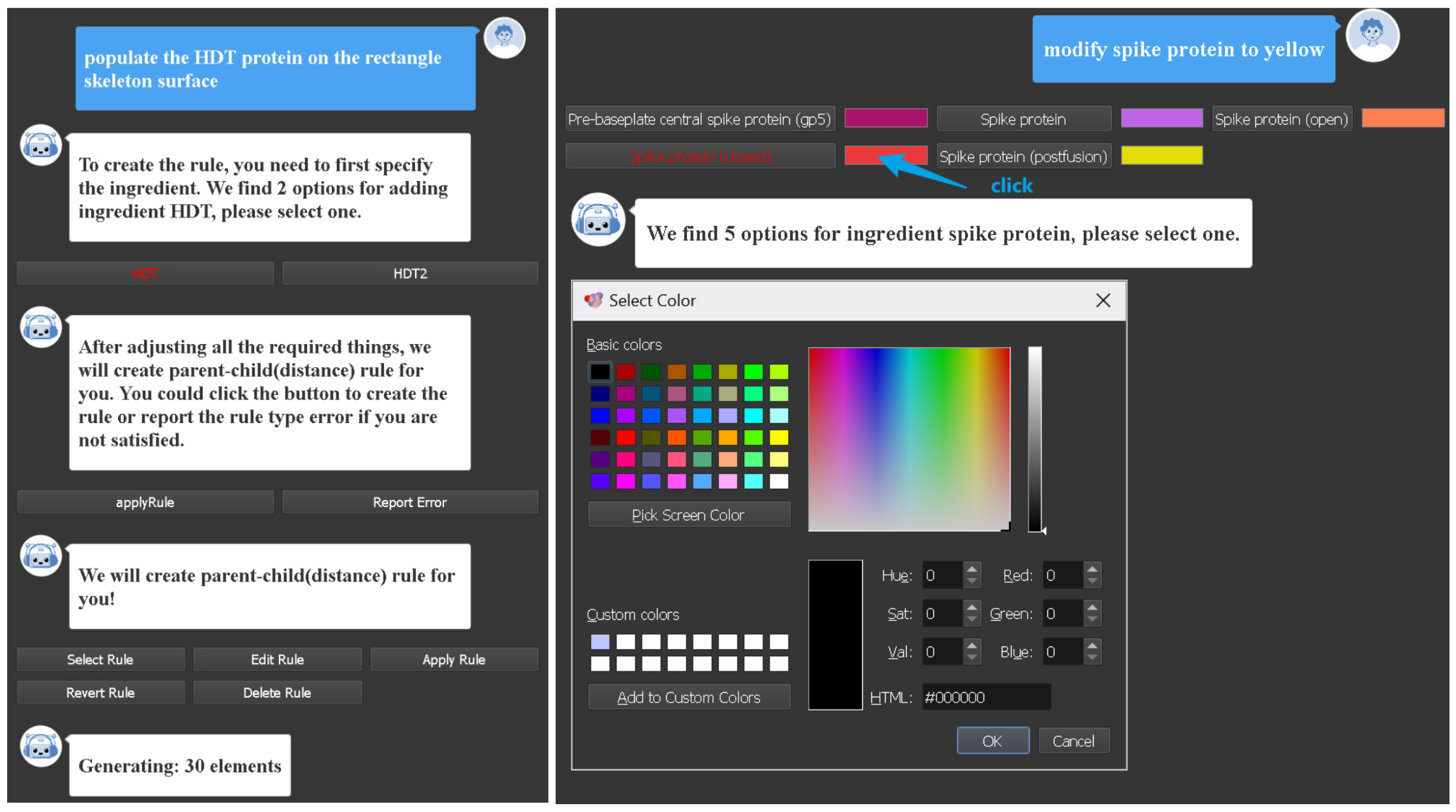}
    \caption{Examples of intent-conditioned widgets. Left: the system generates selection buttons to disambiguate proteins with similar names, as well as control widgets for iterative rule refinement. Right: a visual attribute (color) is first modified through a text instruction and can then be refined using a dynamically generated color picker.}
    \label{fig:placeholder}
\end{figure}

\textbf{Different modeling modes:} Our Chat Modeling supports two complementary modes that reflect how users actually work and the varying planning overhead they are willing to incur. First is the planning mode. Through multi-turn dialogue, the system and user co-author a structured plan. Before execution, the plan is reviewable and editable; this mode suits complex or uncertain tasks where users benefit from receiving the overall strategy from the framework. Second is the step-by-step mode. When users already have a clear plan by themselves but are unfamiliar with the GUI or prefer quick, incremental progress, they can issue each operation through language and receive an immediate result.  We also design an iterative plan execution approach for biology modeling task. Because biological modeling commonly requires fine-tuning at each stage to meet user expectations (DR2), plans are not executed in a single pass. Instead, execution proceeds in a review–refine–advance loop: after each step, users can still issue new language operations or utilize the widgets in the dialog to fine-tune the model. Until satisfied, users then press \emph{Next} to apply the subsequent step. This interleaving of execution with localized parameter tuning preserves control fidelity while avoiding the overhead of re-planning.

\section{Demonstration} \label{sec:Chat Modeling}
\begin{figure}
    \centering
    \includegraphics[width=\linewidth]{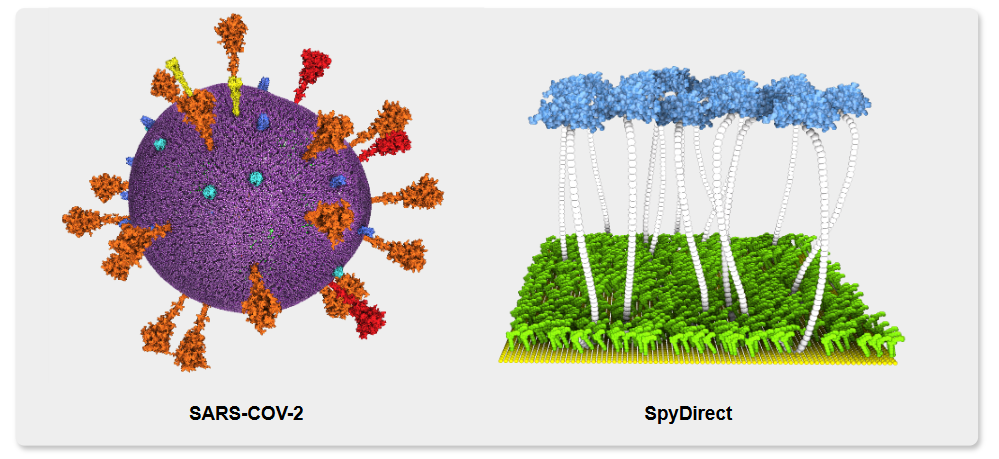}
    \caption{Demonstration models of our two modeling modes.}
    \label{fig:demonstration}
\end{figure}
As a proof of concept, we develop a prototype utilizing our framework, Chat Modeling. This section describes two modeling results using our prototype's different modeling modes. The detailed modeling videos are attached in the supplementary materials. 

\subsection{SARS-COV-2 Model}
Our first case study models a simplified SARS-CoV-2 structure~\cite{nguyen2020modeling}, illustrated in \autoref{fig:demonstration} left. The user adopts an outside-in modeling strategy: first construct the virus’ outer layer and then proceed to the inner layer. The user begins by planning the outer layer. To determine quantities such as the number of surface proteins, the user uploads a research paper on SARS-CoV-2~\cite{yao2020molecular} and asks how many surface proteins should be modeled for the outer layer. Querying the paper, the \emph{Planner} responds: ``A SARS-CoV-2 virion has, on average, $26 \pm 15$ spike proteins randomly distributed on its surface,'' and notes that the paper does not report exact counts for the membrane and envelope proteins. The original wording in the paper is: ``An average of $26 \pm 15$ prefusion S proteins were found to be randomly distributed on each virion.'' With the spike count established, the user requests a final plan for the outer layer, approves it, and begins execution. The plan includes modeling spike proteins in three distinct stages, then modeling the membrane and envelope proteins on the surface, and finally modeling the membrane itself. As the plan is executed, Chat Modeling dynamically generates widgets in the dialogue to support multi-round parameter tuning for each step; the user iteratively adjusts parameters until satisfied and then clicks to advance to the next step.

When preparing to model the membrane, the user realizes that a complete membrane would occlude inner structures. Accordingly, they defer the membrane step, first modeling the inner layer and returning to the membrane afterward. For the inner layer, the user aims to model the RNA strand. Because RNA strand thread through nucleocapsid proteins, the user reuses a previously saved ``RNA strand'' template from the modeling memory. This template models RNA via control points. Using the transfer-knowledge mode, the user specifies that the positions of nucleocapsid proteins should guide the placement of the RNA strands, producing a new plan: first populate nucleocapsid proteins within the skeleton, then connect positions into RNA strands.

After completing the internal RNA strands, the user reclick the previously generated widget for the membrane rule, edits the parameters, and models the full membrane. Notably, the entire workflow is conducted within the dialogue, without hunting through a complex GUI. The intent-conditioned widgets allow rapid refinement at each step; once the inner layer is finished, the user can also revisit earlier widgets to edit prior rules.

\subsection{SpyDirect}
Through the modeling of SARS-CoV-2, we demonstrate structured planning and the reuse of prior modeling templates. The planning mode is best suited for scenarios in which the user does not yet have a concrete modeling plan. When the user already has a clear plan and does not require agent-driven planning, repeatedly invoking a plan–execute loop can be unnecessarily time-consuming. To illustrate how Chat Modeling supports such cases, we reproduce a simplified, proof-of-concept biological structure following Guo \etal~\cite{guo2023spydirect}, illustrated in \autoref{fig:demonstration} right, using the system’s step-by-step mode to construct the model directly and incrementally.

The modeling process begins with a bottom gold layer. This layer consists of gold atoms uniformly arranged across a rectangular skeleton. The instruction provided for this task is, ``Populate the Au atom uniformly on a rectangle skeleton.'' The ``fill'' rule type is identified through the rule extraction operation. After a single round of adjusting the element parameters, the gold layer is successfully generated. The subsequent step involves generating hexanedithiol (HDT) and SpyCatcher instances above the gold layer. Following the user's instructions, Chat Modeling creates two parent-child distance rules. Notably, due to the presence of two similar HDTs in the software, it's necessary for the user to specify the precise HDT variant. This functionality is particularly crucial in biological modeling, where a single protein may have multiple variants across different stages, each with similar names but distinct structural properties. For each parent-child distance rule, the user takes multiple rounds to adjust the rule's parameters, such as element number, the average and standard deviation of the distance, and the tweaking directions. The final step in the modeling process involves creating the protein linkers between HDT and SpyCatcher instances. A protein linker is a short sequence of amino acids used to connect two proteins or domains within a single protein. In our modeling process, we use small balls to represent the amino acids. This protein linker creation process comprises two main steps: initially creating curves between the HDT and SpyCatcher instances and then populating these curves with balls. The user then changes the pivot point of SpyCatcher to adjust the curve's starting point using a specific chain ID and residue ID. Subsequently, curves are formed through the application of a connection rule. The user then proceeds to populate these curves with balls and adjust the number of balls.

\section{Results}\label{sec:numerical analysis}

In this section, we present the quantitative results of our step execution performance on a dataset of task-annotated JSON recipe pairs and discuss the implementation details of Chat Modeling.
\subsection{Quantitative Results} 
To evaluate the effectiveness of our framework, we created a dataset of 200 task-annotated JSON recipe pairs. These tasks were initially gathered from the formative study and were further expanded to encompass a variety of modeling operations, including rule creation, rule parameter adjustment, and visual representation modification. Some tasks involve multiple types of operations; for example, ``populate thirty capsid proteins inside the capsid'' combines rule creation and editing. Detailed tasks are provided in the supplemental material.

We tested our step execution approach (involving recipe generator and recipe interpreter) on two large language models, GPT-4o~\cite{openai_gpt4o_2024} and GPT-4-turbo~\cite{openai2024gpt4}, and compared it to three model settings: zero-shot, few-shot, and our method without recipe correction. We selected accuracy as our evaluation metric. Our correctness criteria are as follows: the format should be consistent with the annotations, and the values of the key-value pairs in the JSON recipe must be accurate. Certain aspects, such as whether an integer is returned as int or float, or minor discrepancies in color values (as some colors don't have standardized RGB values), are acceptable deviations. To ensure a fair quantitative evaluation, we did not involve any modeling memory in any of the settings. The results are summarized in \autoref{tab:comparison}.

\begin{table}[t]
    \centering
    \caption{Numerical results of different methods on our dataset. We report the average accuracy and standard deviation over five runs.}
    \label{tab:comparison}
    \resizebox{\linewidth}{!}{
    \begin{tabular}{lcccc}
        \toprule
        \multirow{2}{*}{\textbf{Method}} 
        & \multicolumn{2}{c}{\textbf{GPT-4o}} 
        & \multicolumn{2}{c}{\textbf{GPT-4-turbo}} \\
        \cmidrule(lr){2-3} \cmidrule(lr){4-5}
        & \textbf{Avg. Acc.} & \textbf{Std.}
        & \textbf{Avg. Acc.} & \textbf{Std.} \\
        \midrule
        Zero-Shot 
        & 0.505 & 0.008 
        & 0.420 & 0.020 \\
        Few-Shot 
        & 0.771 & 0.029 
        & 0.738 & 0.011 \\
        Our Method w/o Recipe Correction 
        & 0.847 & 0.019 
        & 0.810 & 0.018 \\
        \textbf{Our Method} 
        & \textbf{0.918} & \textbf{0.011} 
        & \textbf{0.939} & \textbf{0.014} \\
        \bottomrule
    \end{tabular}
    }
\end{table}

In both zero-shot and few-shot settings, we unified the recipe generator and recipe interpreter within a single prompt, deliberately not employing the recipe correction operation. In the zero-shot setting, no examples are provided, whereas the few-shot setting includes several examples within the prompts. Our findings indicate that the few-shot setting surpasses the zero-shot setting by offering illustrative examples. Furthermore, our method, which excludes recipe correction, outperforms the few-shot setting by completing modeling tasks within several steps. Notably, our approach achieves the highest performance when incorporating recipe correction, thereby enriching the information available for the LLM's reasoning process. These results underscore the effectiveness of our method in completing modeling tasks.

To further evaluate the effectiveness of modeling memory, we conducted a small-scale memory ablation study using 40 sampled tasks from the evaluation set tested on the recipe generation agent. In the baseline condition, the recipe generation agent was prompted in a zero-shot manner without any examples. In the memory condition, we simulated a user-feedback scenario by injecting only one memory example for each task, consisting of the same task query and its corrected recipe output. This setting evaluates whether the system can reuse previously corrected feedback to improve subsequent recipe generation. Using the same correctness criteria as above, GPT-4o improved from 47.5\% accuracy in the zero-shot baseline to 92.5\% with memory, while GPT-4-turbo improved from 45.0\% to 97.5\%. These results provide empirical evidence that the modeling memory can substantially improve recipe generation accuracy after user feedback is incorporated.

\subsection{Implementation Details}

We implement Chat Modeling as a plug-in for the desktop version of MesoCraft, allowing the framework to directly invoke its procedural modeling and visualization pipeline. The prototype uses a chat-based interface connected to an intermediate visualization-action layer, which translates user instructions into structured recipes encoding executable modeling and visualization operations. These recipes are validated before execution to ensure schema consistency and domain-specific constraints, with validation errors returned as textual feedback for correction. The interface also implements the interaction mechanisms described earlier, including intent-conditioned widgets for parameter tuning and representation adjustment, stepwise execution for inspecting intermediate results, and direct mouse interaction for continuous spatial manipulation such as translation and rotation. This implementation enables language, widgets, and direct manipulation to operate within the same MesoCraft workflow.
\section{User Evaluation}
We conducted a two-stage evaluation consisting of an exploratory user study and expert interviews. The exploratory user study focuses on preliminary usability feedback and perceived workload, while the expert interviews assess domain validity, practical usefulness, and real-world applicability from the perspective of biological modeling experts.
\subsection{Exploratory User Study}
We conducted a user study to compare the user experiences of traditional GUI-based modeling and language-based Chat Modeling. A total of six volunteers (three women and three men), denoted as P1–P6, participated in the study. Participants were aged between 24 and 32 years (M = 26.8, SD = 3.4) and were all graduate students in biology-related fields with no prior experience using MesoCraft. Participants were compensated with gift vouchers valued at approximately USD 25. The experiment was conducted by trained staff who had completed data security and bioethics training. The study was approved by the institution’s bioethics committee and was carried out at the institution’s facilities.

\subsubsection{Procedure}
The user study began with a brief introduction to MesoCraft and the Chat Modeling prototype. Participants were then asked to complete a simplified SpyDirect (right side of \autoref{fig:demonstration}) modeling task, described as follows: ``The model includes a gold layer base with hexanedithiol (HDT), above which SpyCatcher proteins are placed and connected to the HDT via linkers.'' To mitigate ordering effects, half of the participants started with the traditional GUI-based modeling interface, while the other half began with the language-based Chat Modeling interface. During the study, the experimenter provided assistance when participants requested help, particularly for unfamiliar GUI operations.

To evaluate user experience and perceived task workload, we employed the System Usability Scale (SUS)~\cite{bangor2008empirical} and the NASA Task Load Index (NASA-TLX)~\cite{hart1988development} for the Chat Modeling prototype. In addition, participants completed a questionnaire assessing Chat Modeling-specific features. Finally, we conducted semi-structured interviews with each participant to collect qualitative insights.

\begin{figure}[h]
    \centering
    \includegraphics[width=0.9\linewidth,height=3.5cm]{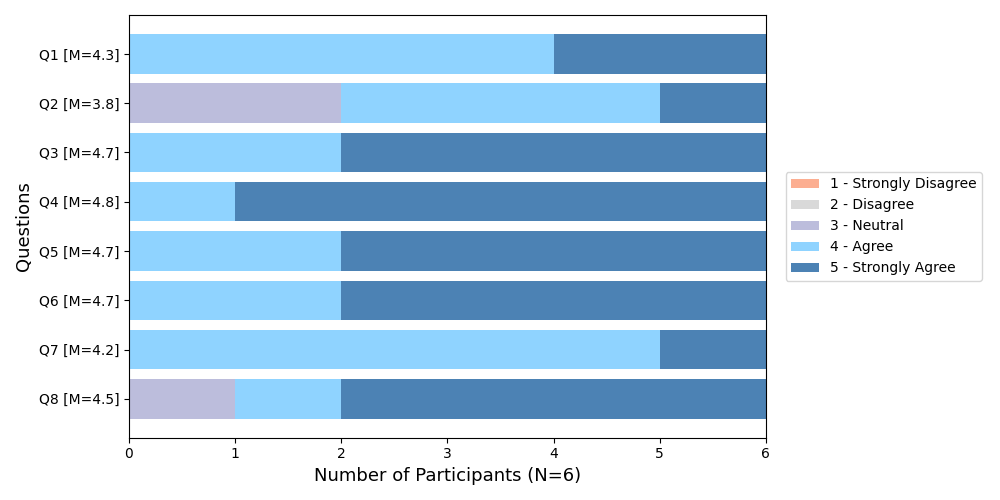}
    \caption{Questionnaire results with a 5-point Likert scale.}
    \label{fig:question}
\end{figure}

\begin{table}[]
    \centering
    \scriptsize
\begin{tabular}{p{8cm}}
\toprule
\textbf{Q1}: It's quick to understand how to perform the modeling tasks. \\
\textbf{Q2}: It's easy to use the system even without prior experience with MesoCraft. \\
\textbf{Q3}: The system allowed users to specify modeling operations at a higher, more intuitive level than traditional GUI-based tools. \\
\textbf{Q4}: Users could express modeling intents easily using natural language in the system. \\
\textbf{Q5}: The system helped users focus more on the modeling goal rather than tool operation. \\
\textbf{Q6}: Users could take control of the modeling process when using the system. \\
\textbf{Q7}: It's easy to correct or refine modeling results in the system. \\
\textbf{Q8}: The system would be useful for users who are new to biological modeling tools. \\

\textbf{I1}: What features are missing in the current system? \\
\textbf{I2}: What features of the current system are most impressive? \\

\bottomrule
\end{tabular}
    \caption{Questionnaire and later interview questions.}
    \label{tab:questionnaire}
\end{table}

\subsubsection{Results}

All six participants successfully completed the study and were included in the analysis. On average, participants completed the tasks in 8.6 min with Chat Modeling and 12.3 min with the GUI-based MesoCraft condition. To assess the overall user experience of the Chat Modeling prototype, we computed the SUS score, which achieved an average of \textbf{80.4} (SD = \textbf{3.68}), indicating high perceived usability. To evaluate perceived task workload, we calculated the weighted NASA-TLX score. Although the original NASA-TLX uses 20-point scales for individual dimensions, we normalized the scores to a 100-point scale for consistency and ease of interpretation, where higher scores indicate lower perceived workload. All dimensions were assigned equal weights when computing the overall score. The average inverted NASA-TLX score was \textbf{81.5} (SD = \textbf{7.0}), suggesting a low workload during task execution. In addition, all participants completed a custom questionnaire targeting Chat Modeling-specific features (see \autoref{tab:questionnaire}) using a 5-point Likert scale. The questionnaire focused on participants’ perceived usability, expressiveness, control, and usefulness of the Chat Modeling prototype. As shown in \autoref{fig:question}, the mean scores for all questions exceeded 4, except Q2, indicating generally positive user feedback toward the system.

\subsubsection{Feedback}

After completing the questionnaire, all participants took part in a follow-up interview consisting of two open-ended questions. Overall, participants expressed a positive attitude toward language-based modeling and found Chat Modeling to be a promising alternative to traditional GUI-driven workflows.

Several participants highlighted the advantages of procedural and language-based modeling over manual operations. In particular, P1 and P6 appreciated the ability to efficiently replicate instances without repetitive manual manipulation, noting that the system ``allows rapid duplication of instances instead of adjusting each one individually.'' P2, P3, and P4 emphasized that natural language interaction lowers the barrier to entry by reducing reliance on detailed GUI knowledge. They noted that even users with little to no prior experience in 3D modeling could quickly begin constructing models using Chat Modeling. P5 noted that an LLM-driven interface is particularly beneficial for providing domain knowledge and assisting with high-level planning during biological modeling tasks. 

At the same time, participants pointed out that Chat Modeling still entails a learning curve, especially for users with no prior exposure to MesoCraft. Several participants remarked that a basic understanding of the underlying modeling concepts and system structure is still required, which helps explain why Q2 did not exceed an average score of 4 in the questionnaire.

Participants also identified several missing features in the current system. P2 and P3 pointed out that the system currently lacks support for fine-grained modifications of individual proteins, such as editing specific residues or adjusting detailed visual properties (e.g., changing the color of a particular amino acid). P1 observed that novice users often issue ambiguous instructions during the planning stage due to limited modeling experience, and suggested that the system could improve usability by proactively prompting users to clarify underspecified commands. Finally, P6 noted that the current system relies on pre-existing MesoCraft skeletons and proteins, and suggested that greater automation, such as automatically querying and downloading relevant PDB files and skeletons, would further enhance the system’s usability and scalability.

\subsection{Expert Interview} \label{sec:expert evaluation}
In this section, we detail the procedure and outcomes of expert evaluations conducted to validate the efficiency of the prototype. 
\subsubsection{Procedure}
To assess the Chat Modeling prototype, we continued our engagement with the three experts initially interviewed in \autoref{sec: formative study}. This choice was driven by multiple considerations. Firstly, their insights are invaluable for verifying whether the prototype meets the expectations set during the formative study and for reflecting on the design requirements. Secondly, their extensive experience in biological structure modeling provides a rare and essential perspective for assessing the prototype against real use cases. Thirdly, their experience guiding new users of MesoCraft equips them with a thorough understanding of the challenges novices face, further enriching their assessment of our prototype.

Our evaluations took the form of semi-structured interviews, with each expert being interviewed individually. The semi-structured format was chosen for its balance of focus and flexibility. This approach enables the investigator to maintain direction on predefined questions while also having the freedom to delve into emerging ideas~\cite{adeoye2021research}. Each session lasted approximately 60 to 90 minutes. Initially, we introduced the prototype to the experts, allowing them a brief overview. We then demonstrate our new modeling workflow from uploading a paper to a final 3D models. Then the experts are invited to freely explore the prototype. This hands-on experience enabled them to provide feedback on the prototype's features and functionalities. Consistent with the nature of semi-structured interviews, the facilitator concluded each session by posing additional questions not previously covered during the interview, ensuring a comprehensive collection of insights and reflections. The detailed semi-structured questions are listed in the supplemental material.

\subsubsection{Outcome}
From the semi-structured interviews, we extracted critical findings about different aspects. These insights, highlighting the application's strengths and areas for improvement, have been instrumental in identifying helpful or missing features. Below, we outline these findings, which collectively inform the direction of future work.

\textbf{Modes:} We summarize feedback from our experts on the two modeling modes, \ie, planning mode and step-by-step mode. Regarding the planning mode, all experts were impressed with this innovative approach to 3D modeling. Nonetheless, they noted that it remains conceptual and requires further improvement in future work. E2 observed, ``It is impressive that it can indicate which of its statements are drawn from the uploaded paper, as this greatly enhances credibility.'' E3 remarked, ``This multimodal input approach expands the possibilities of biological modeling, as the field is characterized by a vast body of literature. A promising future direction is to incorporate multiple papers as input, rather than relying on a single paper as is currently the case.'' For the step-by-step mode, E1 noted, ``This mode is well-suited for users who already know the modeling order and steps of modeling but are unsure how to operate the software. Compared with the planning mode, it also saves the time required for multi-turn dialogues to establish a plan.''

\textbf{Usability:} All experts praised the usability of Chat Modeling, affirming its applicability within their workflows for modeling biological structures. Notably, E1 emphasized the convenience and precision of using natural language for position updates, stating, ``This textual way is more convenient than mouse dragging and offers greater precision because I can specify the residue index.'' E3 remarked, ``The Chat Modeling tool can serve as a copilot, where users can gather inspiration when uncertain about the exact operations.'' We inquired about their opinions on the current response latency, to which they unanimously found it to be entirely acceptable. E1 and E2 highlighted the value of providing initial introductory sentences for new users, commenting, ``The current introductory sentences effectively inform users about how to operate the system. However, it would be beneficial to shorten these introductions once the user is familiar with the system.'' 

\textbf{Interactivity:} Given that Chat Modeling introduces new interactive design, a focal point of our semi-structured interviews was the opinions of the new interactivity.  Regarding the intent-conditioned widgets features, all three experts concurred that ``the combination of mouse clicking and natural language interaction offers a superior modeling experience.'' Furthermore, one expert (E2) remarked, ``Relying solely on natural language interactions may be burdensome for users, especially for actions like applying rules. Therefore, incorporating buttons within the chat box could markedly enhance modeling efficiency.'' E1 emphasized the importance of the selection button's design, noting, ``Users may not always know the exact names of the ingredients due to the presence of variants and complex terminologies. Hence, offering a selection of similar ingredients is a thoughtful design choice. Nonetheless, providing a 3D preview of each ingredient's structure would greatly enrich the user experience.'' For the interactive plan execution feature, E1 noted, ``This interactive execution provides room to adjust the model at each step, since the planning phase does not allow users to see the intermediate results of execution. With interactive execution, I can make real-time adjustments.'' E3 mentioned, ``this interactive execution also serves as a complement to the executed plan, as it allows users to introduce new modeling operations during execution. Before proceeding to the next step, users can still provide additional instructions.''

\textbf{Features:} We inquired with our experts about the aspects of the Chat Modeling prototype they found most valuable, as well as any features or capabilities they felt were missing. All three experts acknowledged the innovation of the modeling memory and its potential to enhance the user experience. E1 suggested, ``Providing feedback through dialogue when I am dissatisfied with an execution result is a valuable feature, as it offers a channel for user feedback. However, the current system requires me to click a button before giving feedback. A more seamless design would enable the system to automatically distinguish whether an instruction is feedback or a modeling operation.'' E3 expressed approval of the high-level parameter design, observing, ``The current parameters in MesoCraft primarily consist of quantitative variables, which do not consider the skeleton's or other elements' properties. Incorporating high-level parameters allows the modeling to reflect real-world properties more accurately, enhancing scientific correctness.'' E2 noted that the ability to reuse previous modeling plans and enable knowledge transfer is highly valuable, as the biological domain contains numerous similar structures. This functionality facilitates the rapid modeling of related structures. Regarding missing features or commands, our experts identified several enhancements that could further improve the tool's functionality. Firstly, they suggested making modifications to visual representations more detailed, such as enabling color adjustments at the residue level. Secondly, they recommended introducing a feature that allows users to upload local ingredients or download ingredients from the internet during their conversation. Lastly, they valued the system's ability to interpret instructions like ``populate more elements,'' turning qualitative statements into quantifiable actions.

Overall, our three experts were pleasantly surprised about the prototype and excited about its possibilities for 3D biological structure modeling. 

\section{Discussion}

\textbf{Biological modeling vs. general 3D modeling:}
Prior autonomous modeling in computer graphics~\cite{kasper2012kit, khalfaoui2013efficient, kriegel2015autonomous}, and more recent text-to-3D generators, primarily pursue perceptual plausibility for everyday objects; in many graphics settings, whether the output is exact is secondary to whether it looks reasonable. Biological visualization has a fundamentally different goal. Here, many components (e.g., proteins) have fixed, experimentally determined shapes, and the task is to compose them under strict geometric and quantitative constraints derived from literature and data. As a result, models must be tightly controlled, auditable, and reproducible, not sampled as ``random but plausible'' content. This distinction motivates our approach: rather than free-form shape synthesis, we emphasize literature-grounded planning and controllable composition of known components, ensuring outputs are scientifically valid, not merely visually convincing.

Beyond biological modeling, we see the most direct applicability of our design in other constraint-driven modeling and visualization tasks where users assemble known components under domain-specific rules, rather than synthesize arbitrary shapes. Examples include  mechanical assembly visualization, materials modeling, and educational simulations of physical systems. In these settings, intent-level control, inspectable plans, and interactive execution can help users refine high-level goals into executable operations while preserving domain constraints. However, the current implementation of Chat Modeling is still specialized for biological modeling: its agents, memory contents, widgets, and execution interface are tied to biological modeling rules and the MesoCraft environment. Applying the framework to other domains would therefore require redefining the domain constraints, replacing the modeling backend, and designing task-specific widgets and validation mechanisms.

\textbf{Failure Cases and Validation}: During the qualitative and expert evaluation, we identified several failure cases that our method cannot address. The main types of these failure cases include mistakenly identifying certain biological concepts, such as cell walls, as skeletons; failing to identify rule parameters during a rule creation operation, such as the parameter specifying ``populate thirty spike proteins on the membrane surface''; and being unable to identify rule types accurately. Additionally, there are operations that our system currently cannot support, which should be addressed in future work. These include operations like ``populate more'' degree representation and ``Place the connexins within the membrane gap.'' The latter task requires precise geometry, spatial information, and collision-handling algorithms to position the elements accurately.

\textbf{Limitations}: Our system represents an early-stage prototype and makes several simplifying design choices. First, it supports interactive refinement during execution but does not yet automatically revise the remaining plan steps after user edits. This static-plan assumption is reasonable for local refinements, such as parameter tuning or minor visual adjustments, but may break down when users skip, reorder, defer, or fundamentally alter steps, since such changes can affect downstream dependencies. Future work should explore partial replanning to distinguish local edits from structural plan changes. Second, our current prototype primarily relies on user inspection and user-reported corrections to identify such errors. This mechanism is useful for capturing explicit feedback, but it does not automatically detect inconsistencies between agents. This places part of the error-detection burden on users, who may not always recognize subtle inter-agent inconsistencies. A promising future direction is to introduce automated validation across agents, such as verifying parameter ranges against domain constraints or rendering-based checks to flag suspicious outputs. Finally, our memory update strategy is based on interaction counts rather than continuous or time-based adaptation, which we leave for future work.

\section{Conclusion}
In this paper, we investigated the potential of the LLM-powered agent framework to enhance scientific visualization and 3D modeling tasks. We introduced a literature-grounded agent framework that assists model planning and execution. Additionally, we propose several interaction designs for plan execution that extend text-only natural language interaction. We also introduce a customized modeling memory for long-term refinement and reusability. Through quantitative evaluation, an exploratory user study, and expert interviews, we gathered insights that underscore the framework’s promise for 3D biological modeling. The feedback highlighted the framework's effectiveness and potential for broader application in scientific modeling.
\section*{Acknowledgments}{
The research was supported by the King Abdullah University of Science and Technology (BAS/1/1680-01-01).
}
\bibliographystyle{IEEEtran}
\bibliography{biography}

@article{nguyen2020modeling,
  author = {Nguyen, Ngan and Strnad, Ond{\v{r}}ej and Klein, Tobias and Luo, Deng and Alharbi, Ruwayda and Wonka, Peter and Maritan, Martina and Mindek, Peter and Autin, Ludovic and Goodsell, David S and others},
  title = {{Modeling in the time of COVID-19: Statistical and rule-based mesoscale models}},
  year = 2020,
  journal = {{IEEE transactions on visualization and computer graphics}},
  publisher = {IEEE},
  volume = 27,
  number = 2,
  pages = {722--732}
}

@article{suris2023vipergpt,
  author = {Sur{\'\i}s, D{\'\i}dac and Menon, Sachit and Vondrick, Carl},
  title = {{Vipergpt: Visual inference via python execution for reasoning}},
  year = 2023,
  journal = {{arXiv preprint arXiv:2303.08128}}
}

@inproceedings{gupta2023visual,
  author = {Gupta, Tanmay and Kembhavi, Aniruddha},
  title = {{Visual programming: Compositional visual reasoning without training}},
  year = 2023,
  booktitle = {{Proceedings of the IEEE/CVF Conference on Computer Vision and Pattern Recognition}},
  pages = {14953--14962}
}

@article{huang2023voxposer,
  author = {Huang, Wenlong and Wang, Chen and Zhang, Ruohan and Li, Yunzhu and Wu, Jiajun and Fei-Fei, Li},
  title = {{Voxposer: Composable 3d value maps for robotic manipulation with language models}},
  year = 2023,
  journal = {{arXiv preprint arXiv:2307.05973}}
}

@article{wang2023prompt,
  author = {Wang, Yen-Jen and Zhang, Bike and Chen, Jianyu and Sreenath, Koushil},
  title = {{Prompt a robot to walk with large language models}},
  year = 2023,
  journal = {{arXiv preprint arXiv:2309.09969}}
}

@article{michel2024object,
  author = {Michel, Oscar and Bhattad, Anand and VanderBilt, Eli and Krishna, Ranjay and Kembhavi, Aniruddha and Gupta, Tanmay},
  title = {{Object 3dit: Language-guided 3d-aware image editing}},
  year = 2024,
  journal = {{Advances in Neural Information Processing Systems}},
  volume = 36
}

@article{sun20233d,
  author = {Sun, Chunyi and Han, Junlin and Deng, Weijian and Wang, Xinlong and Qin, Zishan and Gould, Stephen},
  title = {{3d-gpt: Procedural 3d modeling with large language models}},
  year = 2023,
  journal = {{arXiv preprint arXiv:2310.12945}}
}

@article{feng2024layoutgpt,
  author = {Feng, Weixi and Zhu, Wanrong and Fu, Tsu-jui and Jampani, Varun and Akula, Arjun and He, Xuehai and Basu, Sugato and Wang, Xin Eric and Wang, William Yang},
  title = {{Layoutgpt: Compositional visual planning and generation with large language models}},
  year = 2024,
  journal = {{Advances in Neural Information Processing Systems}},
  volume = 36
}

@article{maddigan2023chat2vis,
  author = {Maddigan, Paula and Susnjak, Teo},
  title = {{Chat2VIS: Fine-Tuning Data Visualisations using Multilingual Natural Language Text and Pre-Trained Large Language Models}},
  year = 2023,
  journal = {{arXiv preprint arXiv:2303.14292}}
}

@article{wang2022towards,
  author = {Wang, Yun and Hou, Zhitao and Shen, Leixian and Wu, Tongshuang and Wang, Jiaqi and Huang, He and Zhang, Haidong and Zhang, Dongmei},
  title = {{Towards natural language-based visualization authoring}},
  year = 2022,
  journal = {{IEEE Transactions on Visualization and Computer Graphics}},
  publisher = {IEEE},
  volume = 29,
  number = 1,
  pages = {1222--1232}
}

@article{guo2023spydirect,
  author = {Guo, Keying and Gr{\"u}nberg, Raik and Ren, Yuxiang and Chang, Tianrui and Wustoni, Shofarul and Strnad, Ondrej and Koklu, Anil and D{\'\i}az-Galicia, Escarlet and Agudelo, Jessica Parrado and Druet, Victor and others},
  title = {{SpyDirect: A Novel Biofunctionalization Method for High Stability and Longevity of Electronic Biosensors}},
  journal = {{Advanced Science}},
  publisher = {Wiley Online Library},
  pages = 2306716
}

@article{narechania2020nl4dv,
  author = {Narechania, Arpit and Srinivasan, Arjun and Stasko, John},
  title = {{NL4DV: A toolkit for generating analytic specifications for data visualization from natural language queries}},
  year = 2020,
  journal = {{IEEE Transactions on Visualization and Computer Graphics}},
  publisher = {IEEE},
  volume = 27,
  number = 2,
  pages = {369--379}
}

@inproceedings{mitra2022facilitating,
  author = {Mitra, Rishab and Narechania, Arpit and Endert, Alex and Stasko, John},
  title = {{Facilitating conversational interaction in natural language interfaces for visualization}},
  year = 2022,
  booktitle = {{2022 IEEE Visualization and Visual Analytics (VIS)}},
  pages = {6--10},
  organization = {IEEE}
}

@inproceedings{raistrick2023infinite,
  author = {Raistrick, Alexander and Lipson, Lahav and Ma, Zeyu and Mei, Lingjie and Wang, Mingzhe and Zuo, Yiming and Kayan, Karhan and Wen, Hongyu and Han, Beining and Wang, Yihan and others},
  title = {{Infinite photorealistic worlds using procedural generation}},
  year = 2023,
  booktitle = {{Proceedings of the IEEE/CVF Conference on Computer Vision and Pattern Recognition}},
  pages = {12630--12641}
}

@misc{BlenderFoundation2023,
  author = {{Blender Foundation}},
  title = {{Blender - a 3D modelling and rendering package}},
  year = 2026,
  note = {Accessed: 2026-01-01},
  howpublished = {\url{https://www.blender.org/}}
}

@misc{AutodeskMaya2023,
  author = {{Autodesk, Inc.}},
  title = {{Maya - 3D Computer Animation, Modeling, Simulation, and Rendering Software}},
  year = 2026,
  note = {Accessed: 2026-01-01},
  howpublished = {\url{https://www.autodesk.ae/products/maya/overview}}
}

@book{prusinkiewicz2012algorithmic,
  author = {Prusinkiewicz, Przemyslaw and Lindenmayer, Aristid},
  title = {{The algorithmic beauty of plants}},
  year = 2012,
  publisher = {Springer Science \& Business Media}
}

@inproceedings{webanck2018procedural,
  author = {Webanck, Antoine and Cortial, Yann and Gu{\'e}rin, Eric and Galin, Eric},
  title = {{Procedural cloudscapes}},
  year = 2018,
  booktitle = {{Computer Graphics Forum}},
  volume = 37,
  number = 2,
  pages = {431--442},
  organization = {Wiley Online Library}
}

@inproceedings{galin2010procedural,
  author = {Galin, Eric and Peytavie, Adrien and Mar{\'e}chal, Nicolas and Gu{\'e}rin, Eric},
  title = {{Procedural generation of roads}},
  year = 2010,
  booktitle = {{Computer Graphics Forum}},
  volume = 29,
  number = 2,
  pages = {429--438},
  organization = {Wiley Online Library}
}

@inproceedings{parish2001procedural,
  author = {Parish, Yoav IH and M{\"u}ller, Pascal},
  title = {{Procedural modeling of cities}},
  year = 2001,
  booktitle = {{Proceedings of the 28th annual conference on Computer graphics and interactive techniques}},
  pages = {301--308}
}

@incollection{muller2006procedural,
  author = {M{\"u}ller, Pascal and Wonka, Peter and Haegler, Simon and Ulmer, Andreas and Van Gool, Luc},
  title = {{Procedural modeling of buildings}},
  year = 2006,
  booktitle = {{ACM SIGGRAPH 2006 Papers}},
  pages = {614--623}
}

@article{schwarz2015advanced,
  author = {Schwarz, Michael and M{\"u}ller, Pascal},
  title = {{Advanced procedural modeling of architecture}},
  year = 2015,
  journal = {{ACM Transactions on Graphics (TOG)}},
  publisher = {ACM New York, NY, USA},
  volume = 34,
  number = 4,
  pages = {1--12}
}

@misc{NanovisG51:online,
  author = {Winfer, John and Syed, Aeliya and Ekers, Paul and Thistle, Leon and Nguyen, Ngan and Strnad, Ondrej and Goodsell, David and Viola, Ivan and Luo, Deng},
  title = {{T4 model}},
  note = {Accessed on 01/01/2026},
  howpublished = {\url{https://www.nanovis.org/T4-model.html}}
}

@misc{NanovisChloroplast:online,
  author = {Mohammed Zia Baig and Ondrej Strnad and Ivan Viola and Deng Luo},
  title = {{Chloroplast Model}},
  note = {Accessed on 01/01/2026},
  howpublished = {\url{https://www.nanovis.org/Chloroplast-model.html}}
}

@article{brown2020language,
  author = {Brown, Tom and Mann, Benjamin and Ryder, Nick and Subbiah, Melanie and Kaplan, Jared D and Dhariwal, Prafulla and Neelakantan, Arvind and Shyam, Pranav and Sastry, Girish and Askell, Amanda and others},
  title = {{Language models are few-shot learners}},
  year = 2020,
  journal = {{Advances in neural information processing systems}},
  volume = 33,
  pages = {1877--1901}
}

@article{adeoye2021research,
  author = {Adeoye-Olatunde, Omolola A and Olenik, Nicole L},
  title = {{Research and scholarly methods: Semi-structured interviews}},
  year = 2021,
  journal = {{Journal of the american college of clinical pharmacy}},
  publisher = {Wiley Online Library},
  volume = 4,
  number = 10,
  pages = {1358--1367}
}

@misc{openai2024gpt4,
  author = {OpenAI},
  title = {{GPT-4 Technical Report}},
  year = 2024,
  eprint = {2303.08774},
  archiveprefix = {arXiv},
  primaryclass = {cs.CL}
}

@article{kasper2012kit,
  author = {Kasper, Alexander and Xue, Zhixing and Dillmann, R{\"u}diger},
  title = {{The kit object models database: An object model database for object recognition, localization and manipulation in service robotics}},
  year = 2012,
  journal = {{The International Journal of Robotics Research}},
  publisher = {SAGE Publications Sage UK: London, England},
  volume = 31,
  number = 8,
  pages = {927--934}
}

@article{khalfaoui2013efficient,
  author = {Khalfaoui, Souhaiel and Seulin, Ralph and Fougerolle, Yohan and Fofi, David},
  title = {{An efficient method for fully automatic 3D digitization of unknown objects}},
  year = 2013,
  journal = {{Computers in Industry}},
  publisher = {Elsevier},
  volume = 64,
  number = 9,
  pages = {1152--1160}
}

@phdthesis{kriegel2015autonomous,
  author = {Kriegel, Simon},
  title = {{Autonomous 3D modeling of unknown objects for active scene exploration}},
  year = 2015,
  school = {Technische Universit{\"a}t M{\"u}nchen (TUM)}
}

@inproceedings{Mohammad_Khalid_2022, series={SA ’22},
   title={CLIP-Mesh: Generating textured meshes from text using pretrained image-text models},
   url={http://dx.doi.org/10.1145/3550469.3555392},
   DOI={10.1145/3550469.3555392},
   booktitle={SIGGRAPH Asia 2022 Conference Papers},
   publisher={ACM},
   author={Mohammad Khalid, Nasir and Xie, Tianhao and Belilovsky, Eugene and Popa, Tiberiu},
   year={2022},
   month=nov, collection={SA ’22} }

@InProceedings{Ma_2023_ICCV,
    author    = {Ma, Yiwei and Zhang, Xiaoqing and Sun, Xiaoshuai and Ji, Jiayi and Wang, Haowei and Jiang, Guannan and Zhuang, Weilin and Ji, Rongrong},
    title     = {X-Mesh: Towards Fast and Accurate Text-driven 3D Stylization via Dynamic Textual Guidance},
    booktitle = {Proceedings of the IEEE/CVF International Conference on Computer Vision (ICCV)},
    month     = {October},
    year      = {2023},
    pages     = {2749-2760}
}

@article{niese2022procedural,
  title={Procedural urban forestry},
  author={Niese, Till and Pirk, S{\"o}ren and Albrecht, Matthias and Benes, Bedrich and Deussen, Oliver},
  journal={ACM Transactions on Graphics (TOG)},
  volume={41},
  number={2},
  pages={1--18},
  year={2022},
  publisher={ACM New York, NY}
}

@article{guo2020inverse,
  title={Inverse procedural modeling of branching structures by inferring L-systems},
  author={Guo, Jianwei and Jiang, Haiyong and Benes, Bedrich and Deussen, Oliver and Zhang, Xiaopeng and Lischinski, Dani and Huang, Hui},
  journal={ACM Transactions on Graphics (TOG)},
  volume={39},
  number={5},
  pages={1--13},
  year={2020},
  publisher={ACM New York, NY, USA}
}

@inproceedings{de2024llmr,
  title={Llmr: Real-time prompting of interactive worlds using large language models},
  author={De La Torre, Fernanda and Fang, Cathy Mengying and Huang, Han and Banburski-Fahey, Andrzej and Amores Fernandez, Judith and Lanier, Jaron},
  booktitle={Proceedings of the CHI Conference on Human Factors in Computing Systems},
  pages={1--22},
  year={2024}
}

@misc{openai_gpt4o_2024,
  title        = {Hello GPT-4o},
  author       = {OpenAI},
  year         = 2024,
  url          = {https://openai.com/index/hello-gpt-4o/},
  note         = {Accessed: 2026-01-01}
}

@misc{lewis2021retrievalaugmentedgenerationknowledgeintensivenlp,
      title={Retrieval-Augmented Generation for Knowledge-Intensive NLP Tasks}, 
      author={Patrick Lewis and Ethan Perez and Aleksandra Piktus and Fabio Petroni and Vladimir Karpukhin and Naman Goyal and Heinrich Küttler and Mike Lewis and Wen-tau Yih and Tim Rocktäschel and Sebastian Riedel and Douwe Kiela},
      year={2021},
      eprint={2005.11401},
      archivePrefix={arXiv},
      primaryClass={cs.CL},
      url={https://arxiv.org/abs/2005.11401}, 
}

@inproceedings{zhong2024memorybank,
  title={Memorybank: Enhancing large language models with long-term memory},
  author={Zhong, Wanjun and Guo, Lianghong and Gao, Qiqi and Ye, He and Wang, Yanlin},
  booktitle={Proceedings of the AAAI Conference on Artificial Intelligence},
  volume={38},
  number={17},
  pages={19724--19731},
  year={2024}
}

@inproceedings{huang2023memory,
  title={Memory sandbox: Transparent and interactive memory management for conversational agents},
  author={Huang, Ziheng and Gutierrez, Sebastian and Kamana, Hemanth and MacNeil, Stephen},
  booktitle={Adjunct Proceedings of the 36th Annual ACM Symposium on User Interface Software and Technology},
  pages={1--3},
  year={2023}
}

@article{chen2021evaluating,
  title={Evaluating large language models trained on code},
  author={Chen, Mark and Tworek, Jerry and Jun, Heewoo and Yuan, Qiming and Pinto, Henrique Ponde De Oliveira and Kaplan, Jared and Edwards, Harri and Burda, Yuri and Joseph, Nicholas and Brockman, Greg and others},
  journal={arXiv preprint arXiv:2107.03374},
  year={2021}
}

@article{zhou2022large,
  title={Large language models are human-level prompt engineers},
  author={Zhou, Yongchao and Muresanu, Andrei Ioan and Han, Ziwen and Paster, Keiran and Pitis, Silviu and Chan, Harris and Ba, Jimmy},
  journal={arXiv preprint arXiv:2211.01910},
  year={2022}
}

@article{yao2020molecular,
  title={Molecular architecture of the SARS-CoV-2 virus},
  author={Yao, Hangping and Song, Yutong and Chen, Yong and Wu, Nanping and Xu, Jialu and Sun, Chujie and Zhang, Jiaxing and Weng, Tianhao and Zhang, Zheyuan and Wu, Zhigang and others},
  journal={Cell},
  volume={183},
  number={3},
  pages={730--738},
  year={2020},
  publisher={Elsevier}
}

@article{bangor2008empirical,
  title={An empirical evaluation of the system usability scale},
  author={Bangor, Aaron and Kortum, Philip T and Miller, James T},
  journal={Intl. Journal of Human--Computer Interaction},
  volume={24},
  number={6},
  pages={574--594},
  year={2008},
  publisher={Taylor \& Francis}
}

@article{hart1988development,
  title={Development of NASA-TLX (Task Load Index): Results of empirical and theoretical research},
  author={Hart, SG},
  journal={Human mental workload/Elsevier},
  year={1988}
}

@article{https://doi.org/10.1111/cgf.13072,
author = {Kozlíková, B. and Krone, M. and Falk, M. and Lindow, N. and Baaden, M. and Baum, D. and Viola, I. and Parulek, J. and Hege, H.-C.},
title = {Visualization of Biomolecular Structures: State of the Art Revisited},
journal = {Computer Graphics Forum},
volume = {36},
number = {8},
pages = {178-204},
doi = {https://doi.org/10.1111/cgf.13072},
url = {https://onlinelibrary.wiley.com/doi/abs/10.1111/cgf.13072},
eprint = {https://onlinelibrary.wiley.com/doi/pdf/10.1111/cgf.13072},
year = {2017}
}

@article{goodsell2020art,
  title={Art and science of the cellular mesoscale},
  author={Goodsell, David S and Olson, Arthur J and Forli, Stefano},
  journal={Trends in Biochemical Sciences},
  volume={45},
  number={6},
  pages={472--483},
  year={2020},
  publisher={Elsevier}
}

@ARTICLE{11037292,
author={Jia, Donggang and Irger, Alexandra and Besancon, Lonni and Strnad, Ondcej and Luo, Deng and Bjorklund, Johanna and Kouyoumdjian, Alexandre and Ynnerman, Anders and Viola, Ivan},
title={{ VOICE: Visual Oracle for Interaction, Conversation, and Explanation }},
year={2025},
volume={31},
number={10},
ISSN={1941-0506},
pages={8828-8845},
doi={10.1109/TVCG.2025.3579956},
url = {https://doi.ieeecomputersociety.org/10.1109/TVCG.2025.3579956},
publisher={IEEE Computer Society},
address={Los Alamitos, CA, USA},
month=oct}
\end{document}